\documentclass[twocolumn,prb,showpacs,superscriptaddress,preprintnumbers,amsmath,amssymb,floatfix]{revtex4}
\usepackage{graphicx}
\usepackage{dcolumn}
\usepackage{bm}
\usepackage{color}
\usepackage{ragged2e}
\usepackage{adjustbox}
\usepackage{color}

\begin{document}

\title{Nature of the magnetism of iridium in the double perovskite Sr$_2$CoIrO$_6$}
\author{S.~Agrestini}
 \email{stefano.agrestini@cells.es}
 \affiliation{Max Planck Institute for Chemical Physics of Solids,
     N\"othnitzerstr. 40, 01187 Dresden, Germany}
 \affiliation{ALBA Synchrotron Light Source, E-08290 Cerdanyola del Vall\`{e}s, Barcelona, Spain}
\author{K.~Chen}
 \affiliation{Synchrotron SOLEIL, L'Orme des Merisiers, Saint-Aubin, 91192 Gif-sur-Yvette, France}
 \affiliation{Institute of Physics II, University of Cologne, Z\"ulpicher Stra{\ss}e 77, 50937 Cologne, Germany}
\author{C.-Y.~Kuo}
  \affiliation{Max Planck Institute for Chemical Physics of Solids,
     N\"othnitzerstr. 40, 01187 Dresden, Germany}
  \affiliation{National Synchrotron Radiation Research Center, 101 Hsin-Ann Road, Hsinchu 30076, Taiwan}
\author{L.~Zhao}
  \affiliation{Max Planck Institute for Chemical Physics of Solids,
     N\"othnitzerstr. 40, 01187 Dresden, Germany}
\author{H.-J.~Lin}
 \affiliation{National Synchrotron Radiation Research Center, 101 Hsin-Ann Road, Hsinchu 30076, Taiwan}
\author{C.-T.~Chen}
 \affiliation{National Synchrotron Radiation Research Center, 101 Hsin-Ann Road, Hsinchu 30076, Taiwan}
\author{A.~Rogalev}
 \affiliation{ESRF-The European Synchrotron, 71 Avenue des Martyrs, 38000 Grenoble, France}
\author{P. Ohresser}
  \affiliation{Synchrotron SOLEIL, L'Orme des Merisiers, Saint-Aubin, 91192 Gif-sur-Yvette, France}
\author{T.-S.~Chan}
 \affiliation{National Synchrotron Radiation Research Center, 101 Hsin-Ann Road, Hsinchu 30076, Taiwan}
\author{S.-C.~Weng}
 \affiliation{National Synchrotron Radiation Research Center, 101 Hsin-Ann Road, Hsinchu 30076, Taiwan}
\author{A.~C.~Komarek}
  \affiliation{Max Planck Institute for Chemical Physics of Solids,
     N\"othnitzerstr. 40, 01187 Dresden, Germany}
\author{K.~Yamaura}
 \affiliation{Research Center for Functional Materials, National Institute for Materials Science, 1-1 Namiki, Tsukuba, Ibaraki 305-0044, Japan}
 \affiliation{Graduate School of Chemical Sciences and Engineering, Hokkaido University, North 10 West 8, Kita-ku, Sapporo, Hokkaido 060-0810, Japan}
\author{M.~W.~Haverkort}
 \affiliation{Institute for theoretical physics, Heidelberg University, Philosophenweg 19, 69120 Heidelberg, Germany}
\author{Z.~Hu}
  \affiliation{Max Planck Institute for Chemical Physics of Solids,
     N\"othnitzerstr. 40, 01187 Dresden, Germany}
\author{L.~H.~Tjeng}
  \affiliation{Max Planck Institute for Chemical Physics of Solids,
     N\"othnitzerstr. 40, 01187 Dresden, Germany}

\date{\today}
\begin{abstract}
We report on our investigation on the magnetism of the iridate double perovskite Sr$_2$CoIrO$_6$, a nominally Ir$^{5+}$  Van Vleck $J_{eff}=0$ system. Using x-ray absorption (XAS) and x-ray magnetic circular dichroism (XMCD) spectroscopy at the Ir-$L_{2,3}$ edges, we found a nearly zero orbital contribution to the magnetic moment and thus an apparent breakdown of the $J_{eff}=0$ ground state. By carrying out also XAS and XMCD experiments at the Co-$L_{2,3}$ edges and by performing detailed full atomic multiplet calculations to simulate all spectra, we discovered that the compound consists of about 90\% Ir$^{5+}$ ($J_{eff}=0$) and Co$^{3+}$ ($S=2$) and 10\% Ir$^{6+}$ ($S=3/2$) and Co$^{2+}$ ($S=3/2$). The magnetic signal of this minority Ir$^{6+}$ component is almost equally strong as that of the main Ir$^{5+}$ component. We infer that there is a competition between the Ir$^{5+}$-Co$^{3+}$ and the Ir$^{6+}$-Co$^{2+}$ configurations in this stoichiometric compound.

\end{abstract}


\maketitle

\section{Introduction }

Recently, correlated oxides with strong spin-orbit coupling (SOC) have attracted a tremendous interest because the associated entanglement of the spin and orbital degrees of freedom may give rise to unexpected exotic electronic states. In the case of iridates with Ir$^{4+}$ (5$d^5$) in octahedral coordination the strong SOC can lead to the so-called $J_{eff}=1/2$ state by splitting the $t_{2g}$ states into a full $j_{eff}=3/2$ band and a half-filled $j_{eff}=1/2$ band \cite{Kim.2008}. This spin-orbit entangled $J_{eff}=1/2$ state renders the Mott insulator behavior observed in many iridium oxides like Sr$_2$IrO$_4$ \cite{Kim.2008,Kim.2009} and has been proposed to provide in honeycomb $d^5$ systems like (Li,Na)$_2$IrO$_3$ \cite{Jackeli.2009,Chaloupka.2010} and RuCl$_3$ \cite{Plumb.2014,Agrestini.2017} the needed prerequisites for the long-sought materialization of the Kitaev model and the emergence of Majorana fermions \cite{Kitaev.2006}.

Applying the same picture of a strong SOC limit to transition metals with $d^4$ configuration, e.g. Ru$^{4+}$, Os$^{4+}$, and Ir$^{5+}$, the $j_{eff}=3/2$ quadruplet is filled with four electrons and the $j_{eff}=1/2$ doublet is completely empty, which leads to a Van Vleck singlet ground state $J_{eff}=0$. In contrast to these expectations, some Ru$^{4+}$ oxides like Ca$_2$RuO$_4$ are known to show an antiferromagnetic ground state \cite{Nakatsuji.1997,Braden.1998}. Recently, theoretical studies have suggested that strong inter-site hopping may lead to superexchange interactions large enough to cause an �exciton condensation�, or more accurately, a condensation of $J_{eff}=1$ triplon excitations, and to drive the antiferromagnetism or ferromagnetism in such nominally Van Vleck $d^4$ systems \cite{Khaliullin.2013,Meetei.2015,Chaloupka.2016}. Later works however, suggested that the interatomic exchange in Ir$^{5+}$ double perovskites might be too weak to overcome the singlet-triplet gap \cite{Pajskr.2016,Svoboda.2017}. Experimentally, on the one hand, Cao et al. \cite{Cao.2014} reported an antiferromagnetic long-range order in the double perovskite Sr$_2$YIrO$_6$ and, to explain the magnetic order, argued that the non-cubic crystal field would cause a suppression of the excitation gap and, as a result, the breakdown of the $J_{eff}=0$ state. On the other hand, a study combining XMCD measurements and full atomic multiplet cluster calculations demonstrated the stability of the Van Vleck singlet state of Ir$^{5+}$, even in presence of strong tetragonal crystal distortions like in Sr$_2$Co$_{0.5}$Ir$_{0.5}$O$_4$ \cite{Agrestini.2018}. A very recent resonant inelastic x-ray scattering study determined the dispersion of the triplet and quintet states in the double perovskites (Ba,Sr)$_2$YIrO$_6$ to be less than 50~meV, i.e. much smaller than the excitation gap, ruling out the possibility of a $J_{eff}=1$ excitonic condensation \cite{Kusch.2018}. The origin of the magnetism reported in the double perovskite Sr$_2$YIrO$_6$ and Ba$_2$YIrO$_6$ is also debated in a number of other papers \cite{Bhowal.2015,Dey.2016,Corredor.2017,Terzic.2017}.

In this context, the magnetism of the Ir$^{5+}$ ion in the double perovskite Sr$_2$CoIrO$_6$ is a very interesting case. In this compound the large difference in cation radii causes the cobalt ions to form a three dimensional alternate arrangement with the iridium ions. Neutron diffraction and susceptibility measurements detected the onset of a long range antiferromagnetic order of the cobalt moments at $T_N \sim 130$~K \cite{Narayanan.2010,Mikhailova.2010}. The Iridium ions, instead, were considered to be paramagnetic. Bond valence sums and band structure calculations predicted Sr$_2$CoIrO$_6$ to have a high spin (HS, $S=2$) Co$^{3+}$ and low spin (LS) Ir$^{5+}$ state \cite{Narayanan.2010}. Surprisingly, a subsequent XMCD study of the La$_{2-x}$Sr$_x$CoIrO$_6$ system reported that the Ir$^{5+}$ has a paramagnetic moment with almost no orbital contribution \cite{Kolchinskaya.2012}, implying that the $J_{eff}=0$ state does not form the ground state. This finding is in contradiction with XMCD studies on other iridates where the Ir$^{5+}$ XMCD signal does indicate the presence of an orbital moment \cite{Agrestini.2018,Laguna.2015}.

Here we address the Sr$_2$CoIrO$_6$ issue by carrying out XAS and XMCD measurements not only on the Ir-$L_{2,3}$ edges but also on the Co-$L_{2,3}$ together with detailed calculations to explain the spectra. Our first objective is to verify whether the valence state of the Ir and Co is 5+ and 3+, respectively, and whether the system is stoichiometric. We then aim to determine what the magnetic ground state is of the Ir$^{5+}$ ions producing possibly such an unusual spectral shape that may indicate the absence of orbital contribution to the paramagnetic moment.

\section{Experimental }

The Sr$_2$CoIrO$_6$ sample was grown from appropriate amounts of SrCO$_3$, Co$_3$O$_4$ and IrO$_2$ that were mixed and ground together. The mixture was pressed into a pellet that was sintered for 22~h at 1180~$^{\circ}$C in air, followed by a final sintering for more than two days in a flow of oxygen ($\sim$ambient pressure).

The XAS at the Co-$L_{2,3}$ edge was recorded in the total electron yield mode at the Dragon beamline of the NSRRC in Taiwan with a photon-energy resolution of 0.25~eV. A single crystal of CoO was measured simultaneously in a separate chamber to obtain relative energy referencing with better than a few meV accuracy at the Co-$L_{3}$ edge (780~eV). The sample pellets were fractured in situ in order to obtain a clean surface. The pressure was below $1\times 10^{-9}$ mbar during the measurements. The XAS at the Co-$K$ and Ir-$L_{3}$ edges were measured in fluorescence yield and transmission modes at the 16A1 and 07A1 beamlines of the NSRRC, respectively. The XMCD spectra at the Co-$L_{2,3}$ edges of Sr$_2$CoIrO$_6$ were collected at the DEIMOS beamline \cite{Ohresser.2014} of the synchrotron SOLEIL in Paris (France) with a photon-energy resolution of 0.2~eV and a degree of circular polarization close to 100\%. The sample was measured at T = 50~K and in a magnetic field of 6~T. The spectra were recorded using the total electron yield method. The sample was also fractured in situ in order to obtain a clean surface. The XMCD measurements at the Ir-$L_{2,3}$ edges were performed at the ID12 beamline \cite{Rogalev.2015} of the European Synchrotron Radiation Facility (ESRF) using the fluorescence yield detection mode. The degree of circular polarization was about 97\%. A self-absorption correction was applied to the Ir-$L_{2,3}$ XAS measured with right and left circular polarized light. Finally the Ir-$L_3/L_2$ edge-jump intensity ratio $I(L_3)/I(L_2)$ was normalized to 2.22 \cite{Henke.1993}. This takes into account the difference in the radial matrix elements of the 2$p_{1/2}$-to-5$d(L_2)$ and 2$p_{3/2}$-to-5$d(L_3)$ transitions. The XMCD spectra were obtained as the direct difference between consecutive x-ray absorption near edge spectroscopy (XANES) scans recorded with opposite helicities of the incoming x-ray beam in 17~T at low temperature of 2~K.

\section{Theoretical calculations }

The configuration-interaction cluster calculations were performed using the Quanty Package \cite{Haverkort.2012,Lu.2014,Haverkort.2014}. The method uses an IrO$_6$ and CoO$_6$ cluster, which includes explicitly the full atomic multiplet interaction, the hybridization of Ir and Co with the ligands, the crystal field acting on the Ir  and Co ions, and the crystal field acting on the ligands. The hybridization strengths and the crystal field parameters were extracted ab initio by DFT calculations carried out using the full-potential local-orbital code FPLO \cite{Koepernik.1999}. The non-cubic crystal field acting on the Ir and Co ions was varied to best fit the experimental XAS and XMCD spectra. The parameters used in the calculations for the Co and Ir ions are listed in Ref. \cite{Coparameters.2019} and \cite{Irparameters.2019}, respectively. Since we are dealing with a polycrystalline sample, we simulated the experimental data by summing two calculated spectra: one for circularly polarized light with the Poynting vector in the xy plane and one with the Poynting vector along the z axis, with a weighting ratio of 2:1. For all simulations we have considered the thermal population of the different states using a Boltzmann distribution.

\section{Experimental Results}
\subsection{Co-$K$ and Ir-$L_3$ XAS}

To check the Co valence we have measured the Co-$K$ edge XAS taken with TFY for La$_2$CoIrO$_6$, Ca$_3$CoRhO$_6$, Sr$_2$CoIrO$_6$ and EuCoO$_3$, as shown in Fig. 1a. Although the spectral features of the Co-$K$ edge are strongly affected by the local crystal structure, the valence state of Co can still be determined \cite{Wong.1984} by the energy position of the steepest slope of the absorption edge. Here we can see that the energy position in Sr$_2$CoIrO$_6$ is the same as that of the Co$^{3+}$ reference sample LaCoO$_3$, and is about ~1.7 eV higher than that of La$_2$CoIrO$_6$ with a Co$^{2+}$ state, which suggests a mainly 3+ valence of the cobalt ions in Sr$_2$CoIrO$_6$. Similar results were obtained previously by A. Kolchinskaya et al. \cite{Kolchinskaya.2012}.

\begin{figure}[t]\centering
\includegraphics[width=\linewidth]{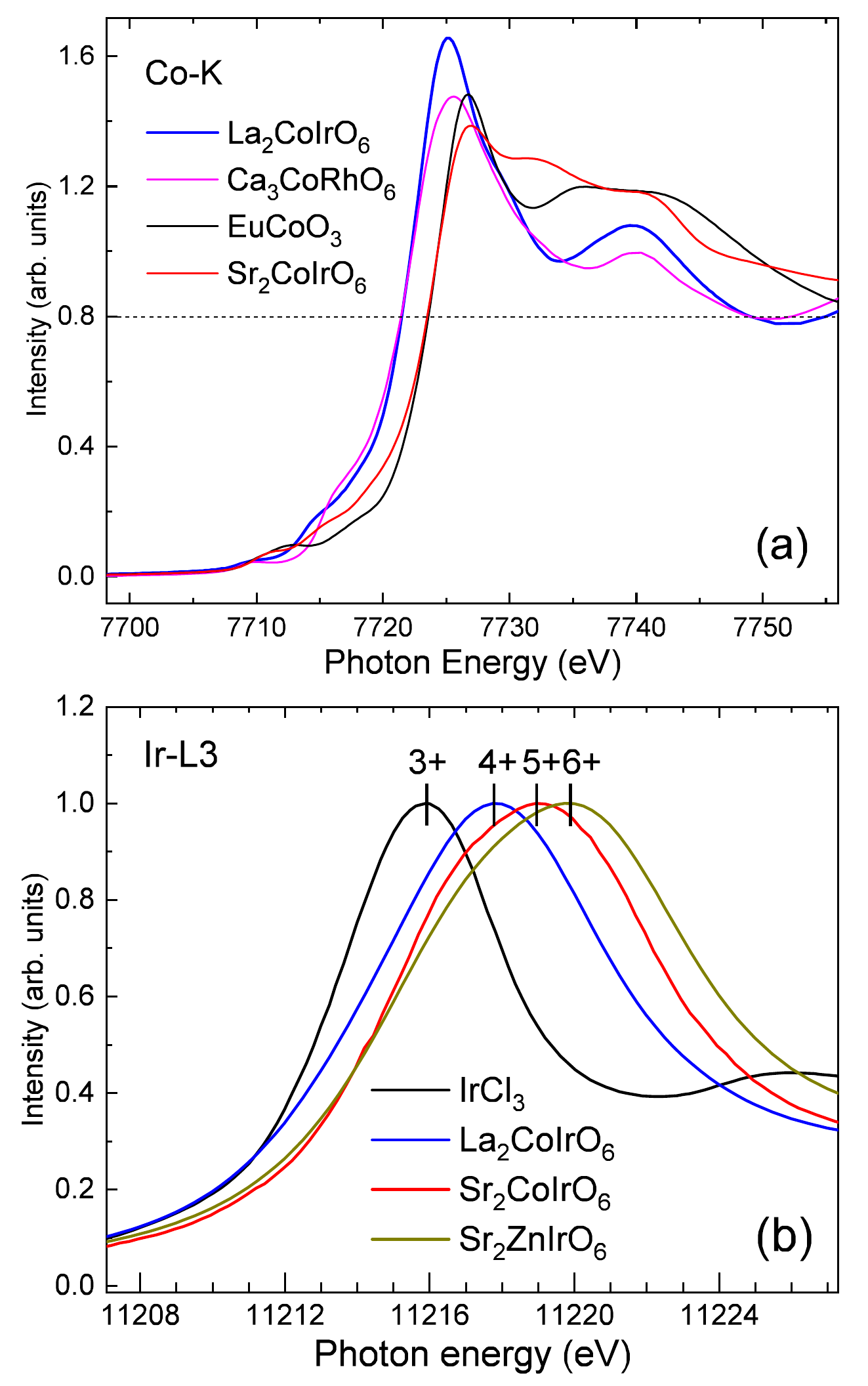}
\caption{(a) The Co-$K$ edge XAS spectra of Sr$_2$CoIrO$_6$ and of La$_2$CoIrO$_6$, and Ca$_3$CoRhO$_6$ as Co$^{2+}$ references, and EuCoO$_3$ as Co$^{3+}$ reference. (b) The Ir-$L_3$ XAS spectra of Sr$_2$CoIrO$_6$ and of IrCl$_3$ as Ir$^{3+}$ reference, La$_2$CoIrO$_6$, and Sr$_2$ZnIr$_6$ as Ir$^{6+}$ reference.}\vspace{-0.2cm}
\label{fig.1}
\end{figure}

Having determined a mainly Co$^{3+}$ state in Sr$_2$CoIrO$_6$, we turn to the Ir-$L_{3}$ XAS spectra to probe the valence of the iridium ions. Fig. 1b reports the Ir-$L_{3}$ XAS spectrum of Sr$_2$CoIrO$_6$ together with the spectra of IrCl$_3$, La$_2$CoIrO$_6$ and Sr$_2$ZnIrO$_6$ as Ir$^{3+}$, Ir$^{4+}$, and Ir$^{6+}$ reference compounds, respectively. It is well known that XAS spectra at the transition metal $L_{2,3}$ edge are highly sensitive to the valence state: an increase of the valence state of the metal ion by one results in a shift of the $L_{2,3}$ XAS spectra by one or more eV toward higher energies, as shown by XAS studies on many oxides \cite{Chen.1990,Mitra.2003,Burnus.2006,Burnus.2008}, including iridium oxides \cite{Laguna.2015,Baroudi.2014,Choy.1995}. This shift is due to a final state effect in the x-ray absorption process. The energy difference between a $d^n$ (e.g. $d^4$ for Ir$^{5+}$) and a $d^{n-1}$ (e.g. $d^3$ for Ir$^{6+}$) configuration is approximately $\Delta$$E = E(2p^6 d^{n-1} \rightarrow 2p^5 d^n) - E(2p^6 d^n \rightarrow 2p^5 d^{n+1}) = U_{pd} - U_{dd} \sim $ 1-2 eV, where $U_{dd}$ is the Coulomb repulsion energy between two $d$ electrons and $U_{pd}$ the one between a $d$ electron and the $2p$ core hole.

One can see that the white line of Sr$_2$CoIrO$_6$ is shifted by $\sim1.3$~eV to higher energies with respect to that of Ir$^{4+}$ in La$_2$CoIrO$_6$, but is shifted by $\sim1$~eV to lower energies with respect to that of Ir$^{6+}$ oxide Sr$_2$ZnIrO$_6$. This energy shift thus indicates a reasonable increase of Ir valence state from 4+ to 5+ and further to 6+ going from La$_2$CoIrO$_6$ to Sr$_2$CoIrO$_6$ and further to Sr$_2$ZnIrO$_6$. A similar energy shift of the white line position was previously observed going from Sr$_3$ZnIr$^{4+}$O$_6$ to Sr$_3$NaIr$^{5+}$O$_6$ and further to Nd$_2$K$_2$Ir$^{6+}$O$_7$ \cite{Mugavero.2009}. Our experimental results are different from the previous study in ref \cite{Kolchinskaya.2012}, where no energy shift of the Ir-$L_{3}$ white-line was observed. Our results are consistent with the above finding of mainly Co$^{3+}$ in Sr$_2$CoIrO$_6$ observed from the Co-$L_{2,3}$, fulfilling the charge balance requirement. It should be noted, however, that the Ir-$L_{3}$ XAS data reported in Fig. 1b cannot exclude the presence of a minor amount of Ir$^{4+}$ or Ir$^{6+}$ ions coexisting with the majority of Ir$^{5+}$ ions.

\subsection{Co-$L_{2,3}$ XAS}

Fig. 2 shows the room temperature Co-$L_{2,3}$ XAS of Sr$_2$CoIrO$_6$ (red line) together with the spectra of EuCoO$_3$ (olive green) as a LS-Co$^{3+}$ reference, Sr$_2$CoRuO$_6$ (black line) as a HS-Co$^{3+}$ reference \cite{Chen.2014}, La$_2$CoIrO$_6$ (blue line) and CoO (green line) as Co$^{2+}$ references. As it can be seen in Fig. 2, the XAS of Sr$_2$CoIrO$_6$ has the centers of gravity of the Co-$L_{2}$ and $L_{3}$ white lines lying at the same energies as those of Sr$_2$CoRuO$_6$ and EuCoO$_3$, and about 1.2~eV higher in energy than those of La$_2$CoIrO$_6$ and CoO. Hence, our experimental Co-$L_{2,3}$ XAS data indicate the cobalt valence state in Sr$_2$CoIrO$_6$ and La$_2$CoIrO$_6$ to be 3+ and 2+, respectively. However, we would like also to point out the presence of a minor prepeak at 778~eV in the spectrum of Sr$_2$CoIrO$_6$. A similar prepeak is also present in the spectrum of Sr$_2$CoRuO$_6$ and was attributed in literature to the presence of a Co$^{2+}$ species \cite{Chen.2014}. By subtracting a 10\% Co$^{2+}$ spectrum from the as measured spectrum of Sr$_2$CoIrO$_6$ one can obtain a XAS spectrum free from any features in the pre-peak region. A two-component scenario with similar amounts of Co$^{2+}$ species was previously reported in literature in thin films of Sr$_2$CoIrO$_6$ \cite{Esser.2018}.

\begin{figure}[t]\centering
\includegraphics[width=\linewidth]{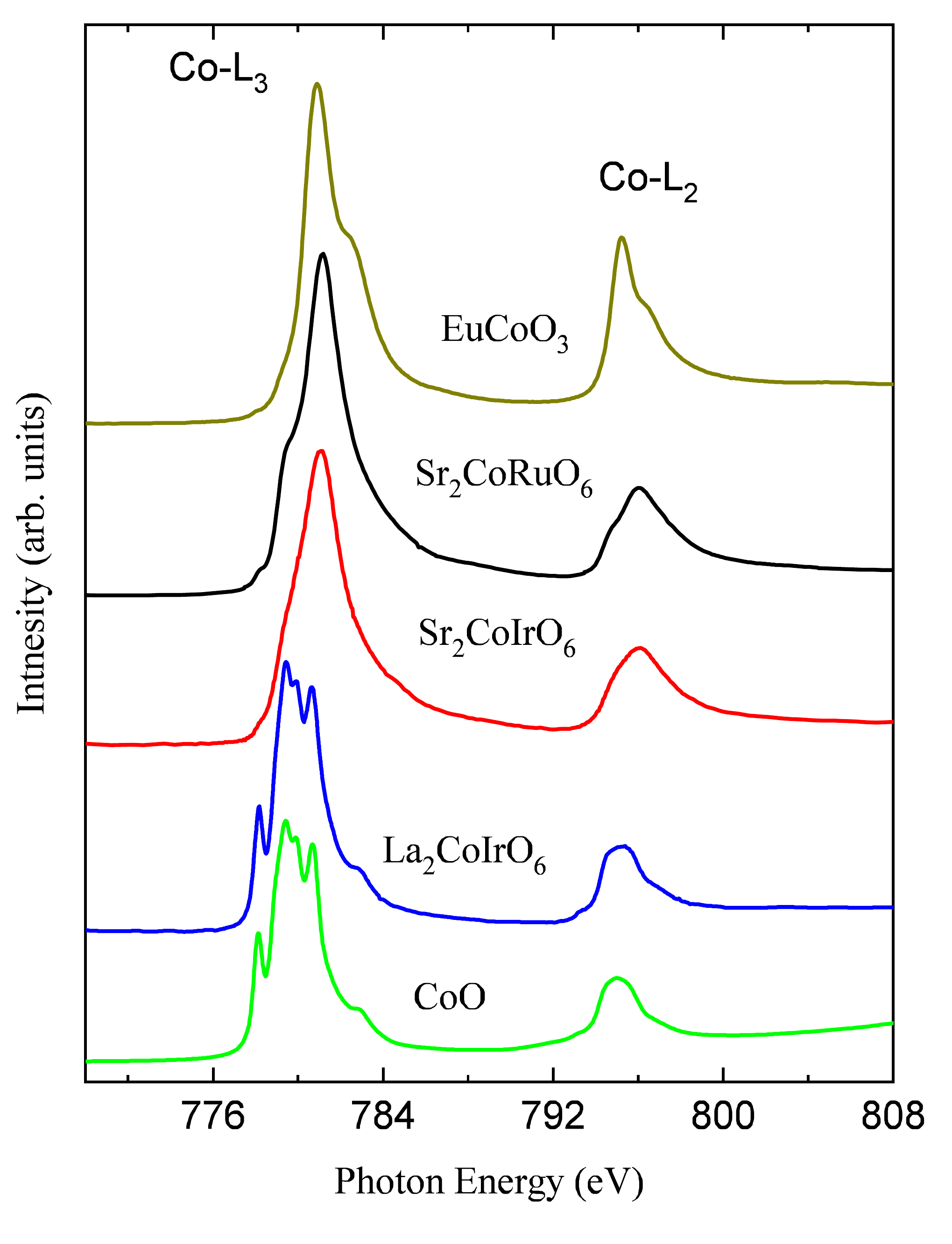}
\caption{ The Co-$L_{2,3}$ absorption of Sr$_2$CoIrO$_6$ (red lines) together with EuCoO$_3$ (olive green), as a LS-Co$^{3+}$ reference, Sr$_2$CoRuO$_6$ (black line) as a HS-Co$^{3+}$ reference, La$_2$CoIrO$_6$ (blue line) and CoO (green line) as a Co$^{2+}$ references.}\vspace{-0.2cm}
\label{fig.2}
\end{figure}

Another unique feature of the $L_{2,3}$ XAS spectra is that the dipole selection rules are very sensitive in determining which of the $2p^53d^{n+1}$ final states can be reached and with what intensity, starting from a particular $2p^63d^n$ initial state ($n=6$ for Co$^{3+}$ and $n=7$ for Co$^{2+}$). This makes the technique extremely sensitive to the symmetry of the initial state, i.e., the spin state and local environment of the Co ions \cite{Hu.2004,Haverkort.2006,Chang.2009,Burnus.2008b,Burnus.2008c,Hollmann.2009}. Despite having the same Co$^{3+}$ valence, the line shape of the Co-$L_{2,3}$ edge spectrum of Sr$_2$CoIrO$_6$ is very different from that of EuCoO$_3$ but in very good agreement with that of Sr$_2$CoRuO$_6$. This shows that the ground state of the Co ions in Sr$_2$CoIrO$_6$ is different from the LS $S=0$ state of EuCoO$_3$ \cite{Hu.2004} and is the same as the HS $S=2$ state of Sr$_2$CoRuO$_6$ \cite{Chen.2014}. The spin only effective magnetic moment of HS Co$^{3+}$ is $\mu_{eff} = 4.9~\mu_B$/f.u. This value in good agreement with the effective magnetic moment $\mu_{eff} = 5.1~\mu_B$/f.u. determined from magnetic susceptibility measurements \cite{Narayanan.2010} assuming a small magnetic moment of the Van Vleck Ir$^{5+}$ ions.

\subsection{Co-$L_{2,3}$ XMCD}

Fig.3 shows the Co-$L_{2,3}$ isotropic XAS and XMCD data (red circles) measured on Sr$_2$CoIrO$_6$ with circular polarized light. The XMCD is defined as the difference between the x-ray absorption spectra taken with the photon spin of the circular polarized light parallel and antiparallel aligned to the magnetic field. In Fig. 3 we have reported also the theoretical Co-$L_{2,3}$ XAS and XMCD spectra for the Co$^{3+}$ in the HS ($S = 2$) configuration (blue lines) as obtained from our full-multiplet configuration-interaction calculations. The HS Co$^{3+}$ simulation can nicely reproduce the line-shape of the measured Co-$L_{2,3}$ XMCD spectrum of Sr$_2$CoIrO$_6$ except for the minor prepeak at 778 eV and the high intense shoulder at 780 eV. These features are related to the XMCD signal of the Co$^{2+}$ ions. If the contribution of the Co$^{2+}$ ions is included through a weighted sum (red lines) of the calculated XMCD (XAS) spectrum of Co$^{3+}$ $S = 2$ and that of Co$^{2+}$ (green lines) the agreement with the experimental XMCD (XAS) spectrum of Sr$_2$CoIrO$_6$ becomes excellent all over the energy range. The simulation provides further evidence for the coexistence of a majority (90\%) of Co$^{3+}$ ions in the $S = 2$ state and a minority (10\%) of Co$^{2+}$ ions in Sr$_2$CoIrO$_6$.

\begin{figure}[t]\centering
\includegraphics[width=\linewidth]{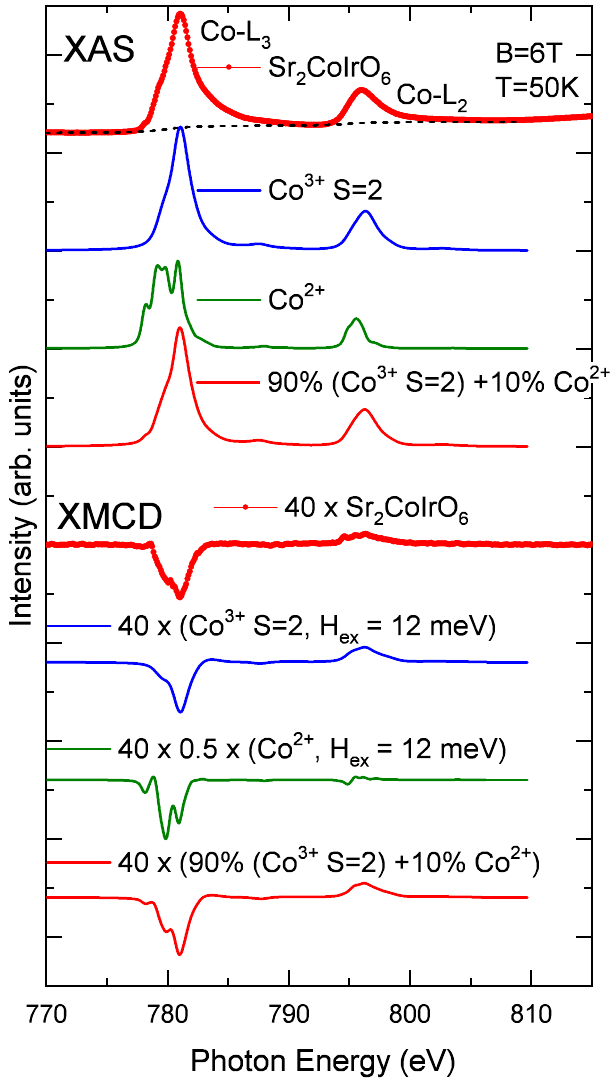}
\caption{ Experimental Co-$L_{2,3}$ XAS and XMCD spectra of Sr$_2$CoIrO$_6$ (red circles) and theoretical simulations: calculated spectra of Co$^{3+}$ $S = 2$ (blue lines), Co$^{2+}$ (green lines) and a weighted sum of calculated spectra of Co$^{3+}$ $S = 2$  and of  Co$^{2+}$ (red lines) for $H_{ex}$ = 12 meV. The simulated XMCD were normalized to the height of the experimental XMCD. The dotted line stands for the edge jump.}\vspace{-0.2cm}
\label{fig.3}
\end{figure}

Important to mention is that the size of the measured Co XMCD is about 11 times smaller than the calculated Co XMCD spectrum if the exchange field $H_{ex}$ is assumed to be zero (paramagnetic case). The small size of the experimental Co XMCD signal is due to the fact the cobalt ions are antiferromagnetically ordered, as revealed by previous neutron diffraction measurements \cite{Narayanan.2010}, and only the canting moment induced by the applied field contributes to the XMCD signal. The size of the experimental XMCD signal was reproduced by using an exchange field of 12 meV, a value that matches nicely with the ordering temperature $T_{N1}$ = 130~K of the cobalt moments \cite{Mikhailova.2010}.

The large difference in intensity of the measured dichroic signal between the $L_{3}$ and $L_{2}$ edges shown in Fig. 3 is a clear signature that the Co ions have a relevant unquenched orbital moment. To be quantitative, we now apply the sum rules for XMCD developed by Thole et al.\cite{Thole.1992} and Carra et al.\cite{Carra.1993}, which provide the orbital to spin ratio:

\begin{equation}\label{eq:ratio}
\frac{L_z}{2S_z+7T_z}=\frac{2}{3}\cdot \frac{\int_{L_{2,3}}(\sigma^+-\sigma^-)dE}{\int_{L_{3}}(\sigma^+-\sigma^-)dE-2\int_{L_{2}}(\sigma^+-\sigma^-)dE}
\end{equation}

\noindent where $T_z$ denotes the intra-atomic magnetic dipole moment. From the experiments we obtain a value
of 0.25 for this quantity. Our configuration-interaction full-multiplet simulation with the weighted sum
of 90\% Co$^{3+}$ and 10\% Co$^{2+}$ provides a value of 0.235, which is in very good agreement
with the experiment. This is fully consistent with the fact that our simulation reproduces very well
the experimental line shapes of the XAS and XMCD spectra as displayed in Fig. 3.

We would like to note that for $3d$ transition metal ions in an octahedral symmetry this term $T_z$ is
a small number \cite{Teramura.1996} and is expected to be a little increased by the local distortion
existing in the present compound. Our configuration interaction full-multiplet calculations indeed found
that the magnetic dipole moment is small compared to the large spin moment the HS Co$^{3+}$ and
Co$^{2+}$: $T_z/S_z = -0.02$. In other words, the above mentioned XMCD sum rule provides in our
case directly the important quantum number of orbital to spin ratio, $L_z/2S_z$.

\subsection{Ir-$L_{2,3}$ XMCD}

The Ir-$L_{2,3}$ XAS and XMCD spectra of Sr$_2$CoIrO$_6$ are reported as red and blue lines, respectively, in Fig. 4, together with the integral of the XMCD signal (green lines). Very important, the Ir-$L_{2}$ and $L_{3}$ XMCD signals have almost equal intensity but opposite sign, which results in a very small integrated intensity (green line) over the Ir-$L_{2,3}$ energy range. Kolchinskaya et al \cite{Kolchinskaya.2012} reported a similar (but not identical) Ir-$L_{2,3}$ XMCD spectrum for Sr$_2$CoIrO$_6$. A vanishing integrated XMCD intensity indicates that the orbital moment of Ir$^{5+}$ in Sr$_2$CoIrO$_6$ is nearly quenched. The spectral lineshape of the present compound is quite different from that of the Ir-$L_{2,3}$ XMCD spectrum of Sr$_2$Co$_{0.5}$Ir$_{0.5}$O$_4$, where the measured dichroic signal at the $L_{3}$ edge is much larger than that at $L_{2}$ edge \cite{Agrestini.2018}. In order to be quantitative we applied the sum rules to our XMCD data and estimated the orbital to spin ratio to be very small and positive: $L_z/(2S_z +7T_z) = 0.03$. This is an order of magnitude smaller than the $L_z/(2S_z +7T_z) = 0.45$ value in Sr$_2$Co$_{0.5}$Ir$_{0.5}$O$_4$ \cite{Agrestini.2018}.

\begin{figure}[t]\centering
\includegraphics[width=\linewidth]{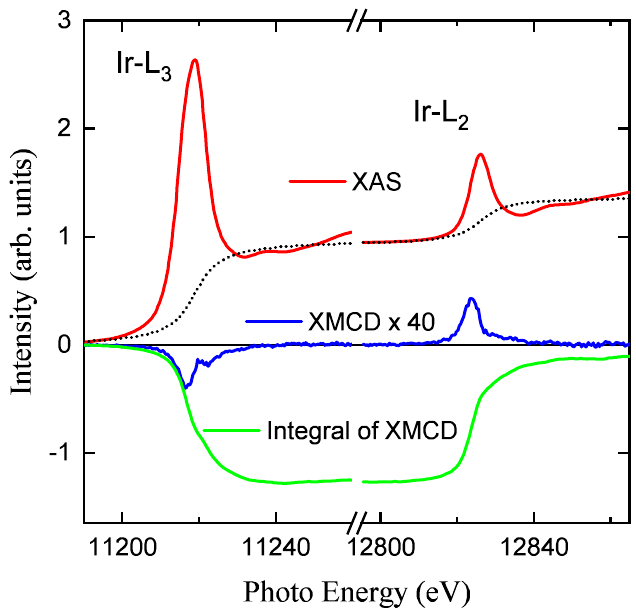}
\caption{ Experimental Ir-$L_{2,3}$ XAS (red line) and XMCD (blue line) spectra measured on Sr$_2$CoIrO$_6$. The green line is the integration over energy of the XMCD signal and the dotted line stands for the edge jump. The spectra were measured at 2~K with a magnetic field of 17~T and using the TFY mode.}\vspace{-0.2cm}
\label{fig.4}
\end{figure}

\section{Discussion}

The Ir-$L_{2,3}$ XMCD spectrum of Sr$_2$CoIrO$_6$ is very different from the usual spectrum measured on other Ir$^{5+}$ oxides, like the layered Sr$_2$Co$_{0.5}$Ir$_{0.5}$O$_4$ \cite{Agrestini.2018} or the double perovskites Sr$_2$MIrO$_6$ with M = Sc, In and Fe \cite{Laguna.2015}. In fact, the typical Ir$^{5+}$ XMCD spectrum exhibits an Ir-$L_{2}$ signal much smaller than the Ir-$L_{3}$ signal. The resulting XMCD integral is large and reflects the presence of a significant orbital moment, as also shown by the application of the sum rules giving an $L_z/2S_z$ ratio ranging from 0.26 (in Sr$_2$FeIrO$_6$) to 0.8 (in Sr$_2$InIrO$_6$). In Sr$_2$Co$_{0.5}$Ir$_{0.5}$O$_4$ \cite{Agrestini.2018} and Sr$_2$ScIrO$_6$ \cite{Laguna.2015} the $L_z/2S_z$ ratio is close to 0.5, i.e. the expected value for a $J_{eff}=0$ ground state. To our knowledge, Sr$_2$CoIrO$_6$ is the only Ir$^{5+}$ oxide displaying a large Ir-$L_{2}$ XMCD signal, with intensity similar to the Ir-$L_{3}$ one, and, hence, having $L_z/2S_z$ close to zero. The question that arises now is what physical mechanism is causing the seemingly vanishing of the orbital moment in the present compound. In order to answer to this question and to determine what is the nature of the ground state of Ir$^{5+}$ ions in Sr$_2$CoIrO$_6$ we have performed configuration-interaction cluster calculations for the Ir-$L_{2,3}$ XAS and XMCD spectra.

\begin{figure}[t]\centering
\includegraphics[width=\linewidth]{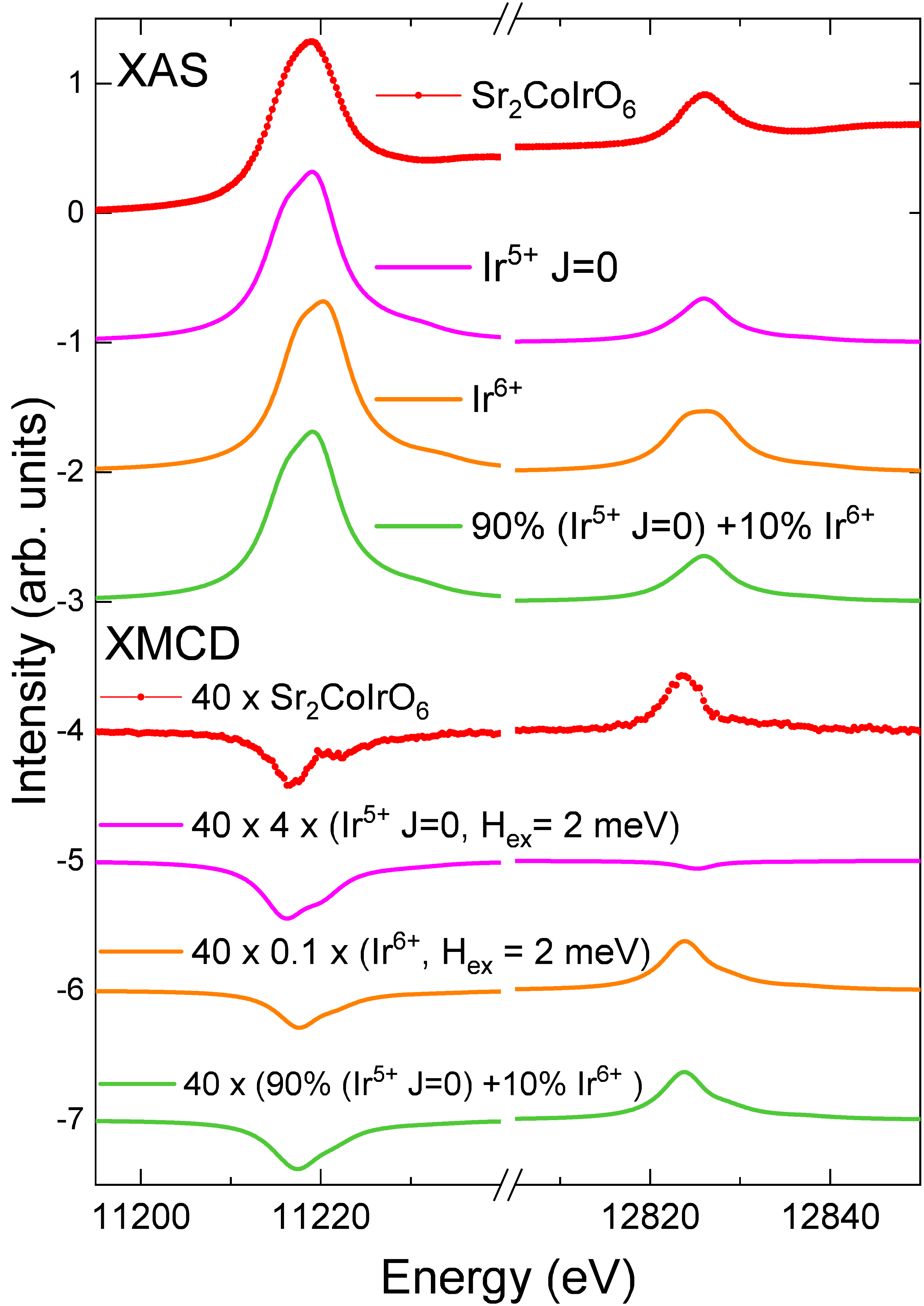}
\caption{ Experimental Ir-$L_{2,3}$ XAS and XMCD spectra of Sr$_2$CoIrO$_6$ (red circles) and theoretical simulations: calculated spectra of Ir$^{5+}$ $J_{eff}=0$  (magenta lines), Ir$^{6+}$ (orange lines)  and a weighted sum of calculated spectra of Ir$^{5+}$ $J_{eff}=0$ and of Ir$^{6+}$ (green lines) for $H_{ex}$ = 2~meV. The simulated XMCD were normalized to the height of the experimental XMCD.  }\vspace{-0.2cm}
\label{fig.5}
\end{figure}

Considering the fact that we have found about 10\% Co$^{2+}$ ions in this formally Co$^{3+}$ system, we investigate the possibility that the measured XMCD signal contains contributions from the presence of Ir$^{4+}$ and/or Ir$^{6+}$ ions in this otherwise Ir$^{5+}$ material. Starting with the Ir$^{4+}$ scenario, we note that the XMCD spectrum of Ir$^{4+}$ oxides is well known to exhibit a very small Ir-$L_{2}$ XMCD signal \cite{Laguna.2015,Laguna.2010,Haskel.2012}. Since this cannot generate the large Ir-$L_{2}$ XMCD signal observed in our Sr$_2$CoIrO$_6$, we can safely rule out the possibility that the XCMD of Sr$_2$CoIrO$_6$ is produced by Ir$^{4+}$ ions. Considering now the Ir$^{6+}$ scenario, we would like to remark that Ir$^{6+}$ ions have a $d^3$ configuration with the spins of three electrons in the $t_{2g}$ shell all parallel to form a $S = 3/2$ spin state. In this situation of half-filled $t_{2g}$ shell the orbital moment is naturally zero or close to. As a consequence, the XMCD spectrum of Ir$^{6+}$ oxides has the $L_{3}$ and $L_{2}$ signals similar in size. This is then a promising scenario to follow.

In Fig. 5, we have plotted the experimental XAS and XMCD spectra together with the simulations for the Ir$^{5+}$ $J_{eff}=0$ (magenta lines) and  Ir$^{6+}$ $S = 3/2$  (orange lines). The parameters for the simulations are listed in Ref. \cite{Irparameters.2019}. We can clearly observe that the calculated XMCD signal at the $L_{2}$ is small for the Ir$^{5+}$ and that it is large for the Ir$^{6+}$, confirming our considerations described in the above paragraph. We now compose a weighted sum of the Ir$^{5+}$ and Ir$^{6+}$ simulations, and the result for a 90:10 ratio is also displayed in Fig. 5 (green lines). This weighted sum can nicely reproduce the line shape of both the experimental XAS and XMCD spectra of Sr$_2$CoIrO$_6$. Hence, the anomalous spectral shape of the Ir-$L_{2,3}$ XMCD of Sr$_2$CoIrO$_6$ can be explained by the presence of 10\%  magnetic Ir$^{6+}$ ions in a matrix of Van Vleck paramagnetic Ir$^{5+}$ ions. Our finding is not inconsistent with a previous diffraction study, where the bond valence sums predicted in Sr$_2$CoIrO$_6$ a partial amount of iridium ions to be in the 6+ valence state \cite{Narayanan.2010}.

In our full multiplet atomic calculations the orbital moment of Ir$^{5+}$ ions is quite large, with an isotropic ratio of $L_z/2S_z = 0.50$ ($L_z/(2S_z +7T_z) = 0.59$). This is the orbital-to-spin moment ratio expected for the $J_{eff} = 0$ ground state \cite{Agrestini.2018}. The calculated orbital moment of the Ir$^{6+}$ ions is very small, as expected for the $S=3/2$ state: $L_z/2S_z = -0.05$ ($L_z/(2S_z +7T_z) = -0.05$). The application of the sum rules to our configuration-interaction full-multiplet simulation of the Ir-$L_{2,3}$ XMCD with the weighted sum of 90\% Ir$^{5+}$ and 10\%  Ir$^{6+}$ provides a value of 0.026, which is in excellent agreement with the experiment.

It is interesting that a 90:10 weighted sum simulation reproduce the experimental spectra quite accurately. The amount of 10\% Ir$^{6+}$ corresponds very well with the presence of 10\% Co$^{2+}$ as we have found earlier. It seems that the two quantities compensate each other, i.e. that the charge balance requirement is fulfilled here. This in turn suggests also that our material is stoichiometric and that there is a competition between the Ir$^{5+}$-Co$^{3+}$ and the Ir$^{6+}$-Co$^{2+}$ configurations in this double perovskite.

It is important to note that the calculated XMCD of the majority Van Vleck Ir$^{5+}$ ions in an applied field of 17 Tesla is roughly half of the size of the needed contribution in the weighted sum to simulate the experimental XMCD spectrum. Such a difference in size can be explained by the presence of a small exchange field of 2~meV acting on the paramagnetic Van Vleck Ir$^{5+}$ ions. The exchange field would be generated by the canting of the antiferromagnetic ordered Co moments, which is induced by the 17~T applied magnetic field, or by the Ir$^{6+}$ ions. A similar exchange field of 2 meV was also used for the calculation of the XMCD of the Ir$^{6+}$ ions. Differently from the paramagnetic Van Vleck Ir$^{5+}$ ions, in our model the Ir$^{6+}$ ions are antiferromagnetically ordered because paramagnetic Ir$^{6+}$ ions would produce a Curie-like divergent susceptibility, which is not observed in the magnetic susceptibility of Sr$_2$CoIrO$_6$ as displayed in Fig. 6. Instead the magnetic susceptibility exhibits a maximum at around 20~K, indicative for the ordering temperature $T_{N2}$ of the Ir$^{6+}$ sublattice.

As final check of our Ir$^{6+}$/ Ir$^{5+}$ two-component scenario for the magnetism of the iridium ions in Sr$_2$CoIrO$_6$, we performed a Curie-Weiss analysis of the magnetic susceptibility. The T-dependent molar magnetic inverse susceptibility $1/(\chi-\chi_0)_{mol}$ (red points) of Sr$_2$CoIrO$_6$ is displayed in Fig. 6. The good linearity of $1/(\chi-\chi_0)_{mol}$  vs. T indicates Curie-Weiss behavior at temperatures above 200~K. Here, we have used $\chi_{0,mol} = 4\times10^{-4}$ emu mol$^{-1}$ Oe$^{-1}$ as also indicated in Fig. 6 (blue line). From the Curie-Weiss fit (green line) we extracted an effective magnetic moment $\mu_{eff}$ of $5.3~\mu_B$ and a Weiss-constant $\theta_W$ of $\sim125$~K. The $|\theta_W|/T_{N1}$ ratio very close to 1 suggests that frustration of the exchange interactions is readily lifted in this compound, like in SrLaNiIrO$_6$ where $|\theta_W|/T_{N1}$ ratio is $\sim$1.2. A very different situation is observed in Ba$_2$BOsO$_6$ (B = Sc, Y, In ) and SrLaCuIrO$_6$, where $|\theta_W|/T_{N1}$ ratios of $\sim$ 6-8 indicate the presence of a large degree of frustration \cite{Feng.2016,Wolff.2019}.

\begin{figure}[t]\centering
\includegraphics[width=\linewidth]{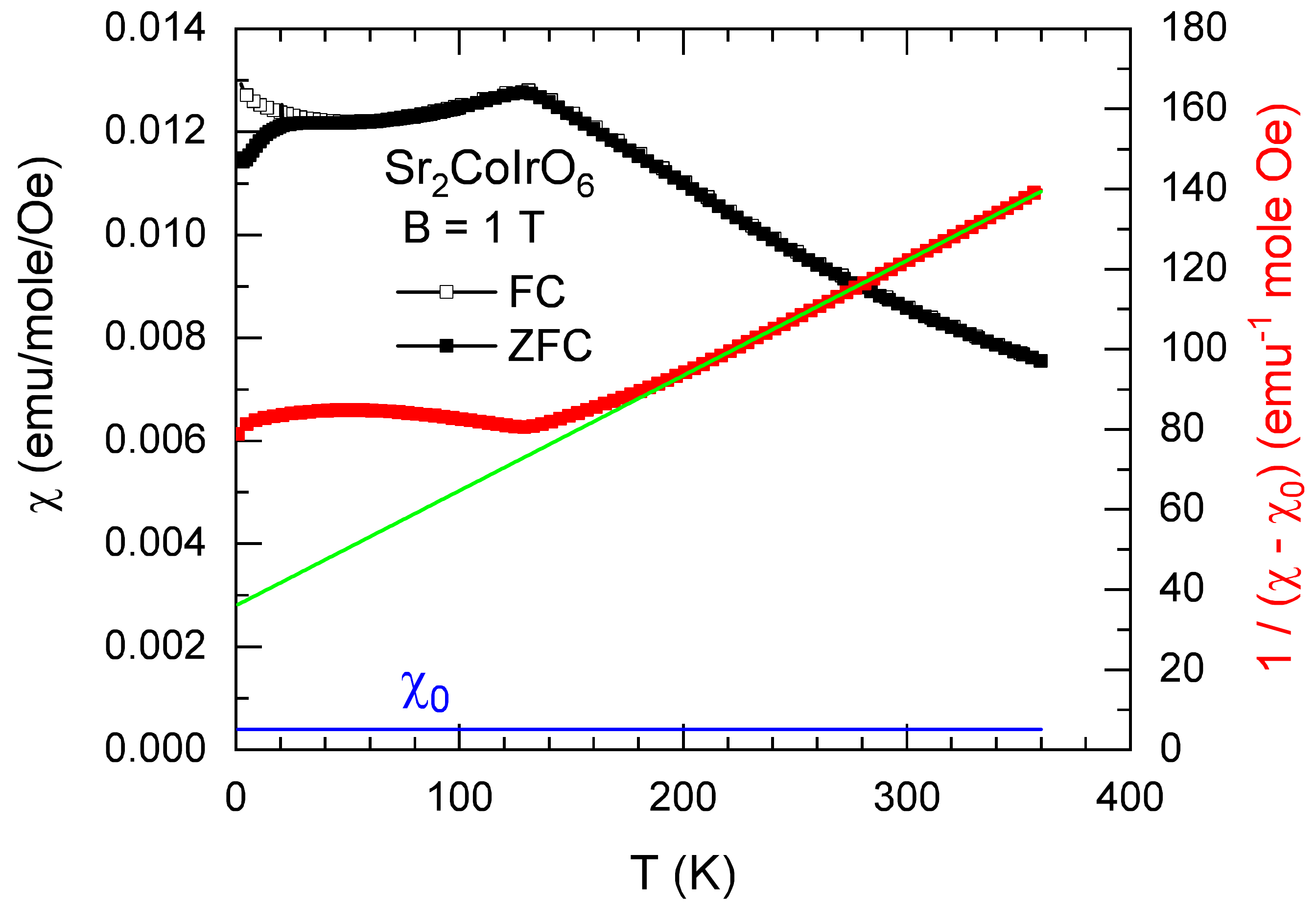}
\caption{ Temperature dependence of the molar and inverse molar magnetic susceptibilities of Sr$_2$CoIrO$_6$ at fields of 1~T in the temperature range of 2-380~K. The Curie-Weiss fit of the inverse susceptibility is shown as green line.  }\vspace{-0.2cm}
\label{fig.6}
\end{figure}

From our full atomic multiplet calculations we found a Van Vleck paramagnetic contribution of $\chi_{0,mol} = 8.2\times10^{-4}$ emu mol$^{-1}$ Oe$^{-1}$ for the $J_{eff}=0$ Ir$^{5+}$ ions ions. This is larger than the experimentally extracted value of $\chi_{0,mol} = 4\times10^{-4}$ emu mol$^{-1}$ Oe$^{-1}$. From the sum of the diamagnetic susceptibilities, obtained from standard charts \cite{Bain.2008}, of the individual ions in the compound we estimate that the temperature independent diamagnetic contribution amounts to about $-1.4 \times10^{-4}$ emu mol$^{-1}$ Oe$^{-1}$. Although we may not be able to fully explain the discrepancy between the calculated and experimental values, it is fair to state that the agreement is quite satisfactory, i.e.  $8.2\times10^{-4}$ vs. $5.4\times10^{-4}$ ($= 4\times10^{-4}$ + $1.4\times10^{-4}$) emu mol$^{-1}$ Oe$^{-1}$. A variety of values in the same range were reported for $\chi_{0,mol}$ of other Ir$^{5+}$ double perovskites : from $10\times10^{-4}$ and $8.7\times10^{-4}$ emu mol$^{-1}$ Oe$^{-1}$ in SrLaCuIrO$_6$ \cite{Wolff.2019} and Sr$_2$YIrO$_6$ \cite{Cao.2014}, respectively, to relatively smaller numbers ($5.8\times10^{-4}$, $5\times10^{-4}$, $3.5\times10^{-4}$ and $3.9\times10^{-4}$ emu mol$^{-1}$ Oe$^{-1}$) in Ba$_2$YIrO$_6$ \cite{Dey.2016}, SrLaNiIrO$_6$, SrLaMgIrO$_6$ and SrLaZnIrO$_6$ \cite{Wolff.2017}.
The total effective magnetic moment of Sr$_2$CoIrO$_6$ as given by our cluster calculations \cite{Coparameters.2019,Irparameters.2019} in the hypothesis of cobalt site 90\% occupied by Co$^{3+}$ and 10\% by Co$^{2+}$, and iridium site 10\% occupied by Ir$^{6+}$ , is $\mu_{eff} (total) = \sqrt{ 0.9\times4.96^2 + 0.1\times5.43^2  + 0.1\times3.07^2 }  = 5.1~\mu_B$.  This value is in good agreement with the value $\mu_{eff}$ of $5.3~\mu_B$ extracted from the Curie-Weiss fit. If on the other hand a pure Ir$^{5+}$ scenario is considered the calculated total $\mu_{eff}$ is reduced to $5.0~\mu_B$. On the base of the above analysis, we can conclude that within the limits of the sensitivity of the magnetic susceptibility the two-component scenario provides a good agreement with the experimental data.

\section{Summary}

To summarize, XAS and XMCD measurements at the Co-$L_{2,3}$ edge demonstrate a Co$^{3+}$ HS state in the double perovskite Sr$_2$CoIrO$_6$. This state is not pure, as our XAS and XMCD also reveal the presence of 10\% cobalt ions in the Co$^{2+}$ state. Our Ir-$L_{2,3}$ edge XAS shows that iridium has mainly the 5+ valence. However, by a comparison of the experimental Ir-$L_{2,3}$ XMCD data with full atomic multiplet calculations we were able to clarify that the signal at the $L_{2}$ edge is mainly due to a contribution from Ir$^{6+}$ ions. Hence, the unusual shape of the XMCD spectrum of Sr$_2$CoIrO$_6$ can be explained with the presence of 10\% of $S = 3/2$ Ir$^{6+}$ ions coexisting with 90\% $J_{eff}=0$ Ir$^{5+}$ ions. The presence of equal amounts of ions with a different valence state in Sr$_2$CoIrO$_6$ is probably driven by the delicate balance between the chemical stability for a Ir$^{5+}$-Co$^{3+}$ configuration versus that for a Ir$^{6+}$-Co$^{2+}$ configuration.

\begin{acknowledgments}
K. C. benefited from support of the Deutsche Forschungsgemeinschaft (DFG) via the Project SE 1441/1-2. The research in Dresden was partially supported by the DFG through SFB 1143 (project-id 247310070) and by the Max Planck-POSTECH-Hsinchu Center for Complex Phase Materials. We gratefully acknowledge SOLEIL, HASYLAB, ESRF and NSRRC for providing us with synchrotron beamtime.
\end{acknowledgments}

\bibliography{Sr2IrCoO6_PRB_2019_05_22}

\begin{thebibliography}{59}
\expandafter\ifx\csname natexlab\endcsname\relax\def\natexlab#1{#1}\fi
\expandafter\ifx\csname bibnamefont\endcsname\relax
  \def\bibnamefont#1{#1}\fi
\expandafter\ifx\csname bibfnamefont\endcsname\relax
  \def\bibfnamefont#1{#1}\fi
\expandafter\ifx\csname citenamefont\endcsname\relax
  \def\citenamefont#1{#1}\fi
\expandafter\ifx\csname url\endcsname\relax
  \def\url#1{\texttt{#1}}\fi
\expandafter\ifx\csname urlprefix\endcsname\relax\def\urlprefix{URL }\fi
\providecommand{\bibinfo}[2]{#2}
\providecommand{\eprint}[2][]{\url{#2}}

\bibitem[{\citenamefont{Kim et~al.}(2008)\citenamefont{Kim, Jin, Moon, Kim,
  Park, Leem, Yu, Noh, Kim, Oh et~al.}}]{Kim.2008}
\bibinfo{author}{\bibfnamefont{B.~J.} \bibnamefont{Kim}},
  \bibinfo{author}{\bibfnamefont{H.}~\bibnamefont{Jin}},
  \bibinfo{author}{\bibfnamefont{S.~J.} \bibnamefont{Moon}},
  \bibinfo{author}{\bibfnamefont{J.-Y.} \bibnamefont{Kim}},
  \bibinfo{author}{\bibfnamefont{B.-G.} \bibnamefont{Park}},
  \bibinfo{author}{\bibfnamefont{C.~S.} \bibnamefont{Leem}},
  \bibinfo{author}{\bibfnamefont{J.}~\bibnamefont{Yu}},
  \bibinfo{author}{\bibfnamefont{T.~W.} \bibnamefont{Noh}},
  \bibinfo{author}{\bibfnamefont{C.}~\bibnamefont{Kim}},
  \bibinfo{author}{\bibfnamefont{S.-J.} \bibnamefont{Oh}},
  \bibnamefont{et~al.}, \bibinfo{journal}{Phys. Rev. Lett.}
  \textbf{\bibinfo{volume}{101}}, \bibinfo{pages}{076402}
  (\bibinfo{year}{2008}),
  \urlprefix\url{https://link.aps.org/doi/10.1103/PhysRevLett.101.076402}.

\bibitem[{\citenamefont{Kim et~al.}(2009)\citenamefont{Kim, Ohsumi, Komesu,
  Sakai, Morita, Takagi, and Arima}}]{Kim.2009}
\bibinfo{author}{\bibfnamefont{B.~J.} \bibnamefont{Kim}},
  \bibinfo{author}{\bibfnamefont{H.}~\bibnamefont{Ohsumi}},
  \bibinfo{author}{\bibfnamefont{T.}~\bibnamefont{Komesu}},
  \bibinfo{author}{\bibfnamefont{S.}~\bibnamefont{Sakai}},
  \bibinfo{author}{\bibfnamefont{T.}~\bibnamefont{Morita}},
  \bibinfo{author}{\bibfnamefont{H.}~\bibnamefont{Takagi}}, \bibnamefont{and}
  \bibinfo{author}{\bibfnamefont{T.}~\bibnamefont{Arima}},
  \bibinfo{journal}{Science} \textbf{\bibinfo{volume}{323}},
  \bibinfo{pages}{1329} (\bibinfo{year}{2009}), ISSN \bibinfo{issn}{0036-8075},
  \eprint{https://science.sciencemag.org/content/323/5919/1329.full.pdf},
  \urlprefix\url{https://science.sciencemag.org/content/323/5919/1329}.

\bibitem[{\citenamefont{Jackeli and Khaliullin}(2009)}]{Jackeli.2009}
\bibinfo{author}{\bibfnamefont{G.}~\bibnamefont{Jackeli}} \bibnamefont{and}
  \bibinfo{author}{\bibfnamefont{G.}~\bibnamefont{Khaliullin}},
  \bibinfo{journal}{Phys. Rev. Lett.} \textbf{\bibinfo{volume}{102}},
  \bibinfo{pages}{017205} (\bibinfo{year}{2009}),
  \urlprefix\url{https://link.aps.org/doi/10.1103/PhysRevLett.102.017205}.

\bibitem[{\citenamefont{Chaloupka et~al.}(2010)\citenamefont{Chaloupka,
  Jackeli, and Khaliullin}}]{Chaloupka.2010}
\bibinfo{author}{\bibfnamefont{J.~c.~v.} \bibnamefont{Chaloupka}},
  \bibinfo{author}{\bibfnamefont{G.}~\bibnamefont{Jackeli}}, \bibnamefont{and}
  \bibinfo{author}{\bibfnamefont{G.}~\bibnamefont{Khaliullin}},
  \bibinfo{journal}{Phys. Rev. Lett.} \textbf{\bibinfo{volume}{105}},
  \bibinfo{pages}{027204} (\bibinfo{year}{2010}),
  \urlprefix\url{https://link.aps.org/doi/10.1103/PhysRevLett.105.027204}.

\bibitem[{\citenamefont{Plumb et~al.}(2014)\citenamefont{Plumb, Clancy,
  Sandilands, Shankar, Hu, Burch, Kee, and Kim}}]{Plumb.2014}
\bibinfo{author}{\bibfnamefont{K.~W.} \bibnamefont{Plumb}},
  \bibinfo{author}{\bibfnamefont{J.~P.} \bibnamefont{Clancy}},
  \bibinfo{author}{\bibfnamefont{L.~J.} \bibnamefont{Sandilands}},
  \bibinfo{author}{\bibfnamefont{V.~V.} \bibnamefont{Shankar}},
  \bibinfo{author}{\bibfnamefont{Y.~F.} \bibnamefont{Hu}},
  \bibinfo{author}{\bibfnamefont{K.~S.} \bibnamefont{Burch}},
  \bibinfo{author}{\bibfnamefont{H.-Y.} \bibnamefont{Kee}}, \bibnamefont{and}
  \bibinfo{author}{\bibfnamefont{Y.-J.} \bibnamefont{Kim}},
  \bibinfo{journal}{Phys. Rev. B} \textbf{\bibinfo{volume}{90}},
  \bibinfo{pages}{041112(R)} (\bibinfo{year}{2014}),
  \urlprefix\url{https://link.aps.org/doi/10.1103/PhysRevB.90.041112}.

\bibitem[{\citenamefont{Agrestini et~al.}(2017)\citenamefont{Agrestini, Kuo,
  Ko, Hu, Kasinathan, Vasili, Herrero-Martin, Valvidares, Pellegrin, Jang
  et~al.}}]{Agrestini.2017}
\bibinfo{author}{\bibfnamefont{S.}~\bibnamefont{Agrestini}},
  \bibinfo{author}{\bibfnamefont{C.-Y.} \bibnamefont{Kuo}},
  \bibinfo{author}{\bibfnamefont{K.-T.} \bibnamefont{Ko}},
  \bibinfo{author}{\bibfnamefont{Z.}~\bibnamefont{Hu}},
  \bibinfo{author}{\bibfnamefont{D.}~\bibnamefont{Kasinathan}},
  \bibinfo{author}{\bibfnamefont{H.~B.} \bibnamefont{Vasili}},
  \bibinfo{author}{\bibfnamefont{J.}~\bibnamefont{Herrero-Martin}},
  \bibinfo{author}{\bibfnamefont{S.~M.} \bibnamefont{Valvidares}},
  \bibinfo{author}{\bibfnamefont{E.}~\bibnamefont{Pellegrin}},
  \bibinfo{author}{\bibfnamefont{L.-Y.} \bibnamefont{Jang}},
  \bibnamefont{et~al.}, \bibinfo{journal}{Phys. Rev. B}
  \textbf{\bibinfo{volume}{96}}, \bibinfo{pages}{161107(R)}
  (\bibinfo{year}{2017}),
  \urlprefix\url{https://link.aps.org/doi/10.1103/PhysRevB.96.161107}.

\bibitem[{\citenamefont{Kitaev}(2006)}]{Kitaev.2006}
\bibinfo{author}{\bibfnamefont{A.}~\bibnamefont{Kitaev}},
  \bibinfo{journal}{Annals of Physics} \textbf{\bibinfo{volume}{321}},
  \bibinfo{pages}{2 } (\bibinfo{year}{2006}), ISSN \bibinfo{issn}{0003-4916},
  \bibinfo{note}{january Special Issue},
  \urlprefix\url{http://www.sciencedirect.com/science/article/pii/S0003491605002381}.

\bibitem[{\citenamefont{Nakatsuji et~al.}(1997)\citenamefont{Nakatsuji, Ikeda,
  and Maeno}}]{Nakatsuji.1997}
\bibinfo{author}{\bibfnamefont{S.}~\bibnamefont{Nakatsuji}},
  \bibinfo{author}{\bibfnamefont{S.-i.} \bibnamefont{Ikeda}}, \bibnamefont{and}
  \bibinfo{author}{\bibfnamefont{Y.}~\bibnamefont{Maeno}},
  \bibinfo{journal}{Journal of the Physical Society of Japan}
  \textbf{\bibinfo{volume}{66}}, \bibinfo{pages}{1868} (\bibinfo{year}{1997}),
  \urlprefix\url{https://doi.org/10.1143/JPSJ.66.1868}.

\bibitem[{\citenamefont{Braden et~al.}(1998)\citenamefont{Braden, Andr\'e,
  Nakatsuji, and Maeno}}]{Braden.1998}
\bibinfo{author}{\bibfnamefont{M.}~\bibnamefont{Braden}},
  \bibinfo{author}{\bibfnamefont{G.}~\bibnamefont{Andr\'e}},
  \bibinfo{author}{\bibfnamefont{S.}~\bibnamefont{Nakatsuji}},
  \bibnamefont{and} \bibinfo{author}{\bibfnamefont{Y.}~\bibnamefont{Maeno}},
  \bibinfo{journal}{Phys. Rev. B} \textbf{\bibinfo{volume}{58}},
  \bibinfo{pages}{847} (\bibinfo{year}{1998}),
  \urlprefix\url{https://link.aps.org/doi/10.1103/PhysRevB.58.847}.

\bibitem[{\citenamefont{Khaliullin}(2013)}]{Khaliullin.2013}
\bibinfo{author}{\bibfnamefont{G.}~\bibnamefont{Khaliullin}},
  \bibinfo{journal}{Phys. Rev. Lett.} \textbf{\bibinfo{volume}{111}},
  \bibinfo{pages}{197201} (\bibinfo{year}{2013}),
  \urlprefix\url{https://link.aps.org/doi/10.1103/PhysRevLett.111.197201}.

\bibitem[{\citenamefont{Meetei et~al.}(2015)\citenamefont{Meetei, Cole,
  Randeria, and Trivedi}}]{Meetei.2015}
\bibinfo{author}{\bibfnamefont{O.~N.} \bibnamefont{Meetei}},
  \bibinfo{author}{\bibfnamefont{W.~S.} \bibnamefont{Cole}},
  \bibinfo{author}{\bibfnamefont{M.}~\bibnamefont{Randeria}}, \bibnamefont{and}
  \bibinfo{author}{\bibfnamefont{N.}~\bibnamefont{Trivedi}},
  \bibinfo{journal}{Phys. Rev. B} \textbf{\bibinfo{volume}{91}},
  \bibinfo{pages}{054412} (\bibinfo{year}{2015}),
  \urlprefix\url{https://link.aps.org/doi/10.1103/PhysRevB.91.054412}.

\bibitem[{\citenamefont{Chaloupka and Khaliullin}(2016)}]{Chaloupka.2016}
\bibinfo{author}{\bibfnamefont{J.~c.~v.} \bibnamefont{Chaloupka}}
  \bibnamefont{and}
  \bibinfo{author}{\bibfnamefont{G.}~\bibnamefont{Khaliullin}},
  \bibinfo{journal}{Phys. Rev. Lett.} \textbf{\bibinfo{volume}{116}},
  \bibinfo{pages}{017203} (\bibinfo{year}{2016}),
  \urlprefix\url{https://link.aps.org/doi/10.1103/PhysRevLett.116.017203}.

\bibitem[{\citenamefont{Pajskr et~al.}(2016)\citenamefont{Pajskr, Nov\'ak,
  Pokorn\'y, Koloren\ifmmode~\check{c}\else \v{c}\fi{}, Arita, and
  Kune\ifmmode~\check{s}\else \v{s}\fi{}}}]{Pajskr.2016}
\bibinfo{author}{\bibfnamefont{K.}~\bibnamefont{Pajskr}},
  \bibinfo{author}{\bibfnamefont{P.}~\bibnamefont{Nov\'ak}},
  \bibinfo{author}{\bibfnamefont{V.}~\bibnamefont{Pokorn\'y}},
  \bibinfo{author}{\bibfnamefont{J.}~\bibnamefont{Koloren\ifmmode~\check{c}\else
  \v{c}\fi{}}}, \bibinfo{author}{\bibfnamefont{R.}~\bibnamefont{Arita}},
  \bibnamefont{and}
  \bibinfo{author}{\bibfnamefont{J.}~\bibnamefont{Kune\ifmmode~\check{s}\else
  \v{s}\fi{}}}, \bibinfo{journal}{Phys. Rev. B} \textbf{\bibinfo{volume}{93}},
  \bibinfo{pages}{035129} (\bibinfo{year}{2016}),
  \urlprefix\url{https://link.aps.org/doi/10.1103/PhysRevB.93.035129}.

\bibitem[{\citenamefont{Svoboda et~al.}(2017)\citenamefont{Svoboda, Randeria,
  and Trivedi}}]{Svoboda.2017}
\bibinfo{author}{\bibfnamefont{C.}~\bibnamefont{Svoboda}},
  \bibinfo{author}{\bibfnamefont{M.}~\bibnamefont{Randeria}}, \bibnamefont{and}
  \bibinfo{author}{\bibfnamefont{N.}~\bibnamefont{Trivedi}},
  \bibinfo{journal}{Phys. Rev. B} \textbf{\bibinfo{volume}{95}},
  \bibinfo{pages}{014409} (\bibinfo{year}{2017}),
  \urlprefix\url{https://link.aps.org/doi/10.1103/PhysRevB.95.014409}.

\bibitem[{\citenamefont{Cao et~al.}(2014)\citenamefont{Cao, Qi, Li, Terzic,
  Yuan, DeLong, Murthy, and Kaul}}]{Cao.2014}
\bibinfo{author}{\bibfnamefont{G.}~\bibnamefont{Cao}},
  \bibinfo{author}{\bibfnamefont{T.~F.} \bibnamefont{Qi}},
  \bibinfo{author}{\bibfnamefont{L.}~\bibnamefont{Li}},
  \bibinfo{author}{\bibfnamefont{J.}~\bibnamefont{Terzic}},
  \bibinfo{author}{\bibfnamefont{S.~J.} \bibnamefont{Yuan}},
  \bibinfo{author}{\bibfnamefont{L.~E.} \bibnamefont{DeLong}},
  \bibinfo{author}{\bibfnamefont{G.}~\bibnamefont{Murthy}}, \bibnamefont{and}
  \bibinfo{author}{\bibfnamefont{R.~K.} \bibnamefont{Kaul}},
  \bibinfo{journal}{Phys. Rev. Lett.} \textbf{\bibinfo{volume}{112}},
  \bibinfo{pages}{056402} (\bibinfo{year}{2014}),
  \urlprefix\url{https://link.aps.org/doi/10.1103/PhysRevLett.112.056402}.

\bibitem[{\citenamefont{Agrestini et~al.}(2018)\citenamefont{Agrestini, Kuo,
  Chen, Utsumi, Mikhailova, Rogalev, Wilhelm, F\"orster, Matsumoto, Takayama
  et~al.}}]{Agrestini.2018}
\bibinfo{author}{\bibfnamefont{S.}~\bibnamefont{Agrestini}},
  \bibinfo{author}{\bibfnamefont{C.-Y.} \bibnamefont{Kuo}},
  \bibinfo{author}{\bibfnamefont{K.}~\bibnamefont{Chen}},
  \bibinfo{author}{\bibfnamefont{Y.}~\bibnamefont{Utsumi}},
  \bibinfo{author}{\bibfnamefont{D.}~\bibnamefont{Mikhailova}},
  \bibinfo{author}{\bibfnamefont{A.}~\bibnamefont{Rogalev}},
  \bibinfo{author}{\bibfnamefont{F.}~\bibnamefont{Wilhelm}},
  \bibinfo{author}{\bibfnamefont{T.}~\bibnamefont{F\"orster}},
  \bibinfo{author}{\bibfnamefont{A.}~\bibnamefont{Matsumoto}},
  \bibinfo{author}{\bibfnamefont{T.}~\bibnamefont{Takayama}},
  \bibnamefont{et~al.}, \bibinfo{journal}{Phys. Rev. B}
  \textbf{\bibinfo{volume}{97}}, \bibinfo{pages}{214436}
  (\bibinfo{year}{2018}),
  \urlprefix\url{https://link.aps.org/doi/10.1103/PhysRevB.97.214436}.

\bibitem[{\citenamefont{Kusch et~al.}(2018)\citenamefont{Kusch, Katukuri,
  Bogdanov, B\"uchner, Dey, Efremov, Hamann-Borrero, Kim, Krisch, Maljuk
  et~al.}}]{Kusch.2018}
\bibinfo{author}{\bibfnamefont{M.}~\bibnamefont{Kusch}},
  \bibinfo{author}{\bibfnamefont{V.~M.} \bibnamefont{Katukuri}},
  \bibinfo{author}{\bibfnamefont{N.~A.} \bibnamefont{Bogdanov}},
  \bibinfo{author}{\bibfnamefont{B.}~\bibnamefont{B\"uchner}},
  \bibinfo{author}{\bibfnamefont{T.}~\bibnamefont{Dey}},
  \bibinfo{author}{\bibfnamefont{D.~V.} \bibnamefont{Efremov}},
  \bibinfo{author}{\bibfnamefont{J.~E.} \bibnamefont{Hamann-Borrero}},
  \bibinfo{author}{\bibfnamefont{B.~H.} \bibnamefont{Kim}},
  \bibinfo{author}{\bibfnamefont{M.}~\bibnamefont{Krisch}},
  \bibinfo{author}{\bibfnamefont{A.}~\bibnamefont{Maljuk}},
  \bibnamefont{et~al.}, \bibinfo{journal}{Phys. Rev. B}
  \textbf{\bibinfo{volume}{97}}, \bibinfo{pages}{064421}
  (\bibinfo{year}{2018}),
  \urlprefix\url{https://link.aps.org/doi/10.1103/PhysRevB.97.064421}.

\bibitem[{\citenamefont{Bhowal et~al.}(2015)\citenamefont{Bhowal, Baidya,
  Dasgupta, and Saha-Dasgupta}}]{Bhowal.2015}
\bibinfo{author}{\bibfnamefont{S.}~\bibnamefont{Bhowal}},
  \bibinfo{author}{\bibfnamefont{S.}~\bibnamefont{Baidya}},
  \bibinfo{author}{\bibfnamefont{I.}~\bibnamefont{Dasgupta}}, \bibnamefont{and}
  \bibinfo{author}{\bibfnamefont{T.}~\bibnamefont{Saha-Dasgupta}},
  \bibinfo{journal}{Phys. Rev. B} \textbf{\bibinfo{volume}{92}},
  \bibinfo{pages}{121113(R)} (\bibinfo{year}{2015}),
  \urlprefix\url{https://link.aps.org/doi/10.1103/PhysRevB.92.121113}.

\bibitem[{\citenamefont{Dey et~al.}(2016)\citenamefont{Dey, Maljuk, Efremov,
  Kataeva, Gass, Blum, Steckel, Gruner, Ritschel, Wolter et~al.}}]{Dey.2016}
\bibinfo{author}{\bibfnamefont{T.}~\bibnamefont{Dey}},
  \bibinfo{author}{\bibfnamefont{A.}~\bibnamefont{Maljuk}},
  \bibinfo{author}{\bibfnamefont{D.~V.} \bibnamefont{Efremov}},
  \bibinfo{author}{\bibfnamefont{O.}~\bibnamefont{Kataeva}},
  \bibinfo{author}{\bibfnamefont{S.}~\bibnamefont{Gass}},
  \bibinfo{author}{\bibfnamefont{C.~G.~F.} \bibnamefont{Blum}},
  \bibinfo{author}{\bibfnamefont{F.}~\bibnamefont{Steckel}},
  \bibinfo{author}{\bibfnamefont{D.}~\bibnamefont{Gruner}},
  \bibinfo{author}{\bibfnamefont{T.}~\bibnamefont{Ritschel}},
  \bibinfo{author}{\bibfnamefont{A.~U.~B.} \bibnamefont{Wolter}},
  \bibnamefont{et~al.}, \bibinfo{journal}{Phys. Rev. B}
  \textbf{\bibinfo{volume}{93}}, \bibinfo{pages}{014434}
  (\bibinfo{year}{2016}),
  \urlprefix\url{https://link.aps.org/doi/10.1103/PhysRevB.93.014434}.

\bibitem[{\citenamefont{Corredor et~al.}(2017)\citenamefont{Corredor,
  Aslan-Cansever, Sturza, Manna, Maljuk, Gass, Dey, Wolter, Kataeva, Zimmermann
  et~al.}}]{Corredor.2017}
\bibinfo{author}{\bibfnamefont{L.~T.} \bibnamefont{Corredor}},
  \bibinfo{author}{\bibfnamefont{G.}~\bibnamefont{Aslan-Cansever}},
  \bibinfo{author}{\bibfnamefont{M.}~\bibnamefont{Sturza}},
  \bibinfo{author}{\bibfnamefont{K.}~\bibnamefont{Manna}},
  \bibinfo{author}{\bibfnamefont{A.}~\bibnamefont{Maljuk}},
  \bibinfo{author}{\bibfnamefont{S.}~\bibnamefont{Gass}},
  \bibinfo{author}{\bibfnamefont{T.}~\bibnamefont{Dey}},
  \bibinfo{author}{\bibfnamefont{A.~U.~B.} \bibnamefont{Wolter}},
  \bibinfo{author}{\bibfnamefont{O.}~\bibnamefont{Kataeva}},
  \bibinfo{author}{\bibfnamefont{A.}~\bibnamefont{Zimmermann}},
  \bibnamefont{et~al.}, \bibinfo{journal}{Phys. Rev. B}
  \textbf{\bibinfo{volume}{95}}, \bibinfo{pages}{064418}
  (\bibinfo{year}{2017}),
  \urlprefix\url{https://link.aps.org/doi/10.1103/PhysRevB.95.064418}.

\bibitem[{\citenamefont{Terzic et~al.}(2017)\citenamefont{Terzic, Zheng, Ye,
  Zhao, Schlottmann, De~Long, Yuan, and Cao}}]{Terzic.2017}
\bibinfo{author}{\bibfnamefont{J.}~\bibnamefont{Terzic}},
  \bibinfo{author}{\bibfnamefont{H.}~\bibnamefont{Zheng}},
  \bibinfo{author}{\bibfnamefont{F.}~\bibnamefont{Ye}},
  \bibinfo{author}{\bibfnamefont{H.~D.} \bibnamefont{Zhao}},
  \bibinfo{author}{\bibfnamefont{P.}~\bibnamefont{Schlottmann}},
  \bibinfo{author}{\bibfnamefont{L.~E.} \bibnamefont{De~Long}},
  \bibinfo{author}{\bibfnamefont{S.~J.} \bibnamefont{Yuan}}, \bibnamefont{and}
  \bibinfo{author}{\bibfnamefont{G.}~\bibnamefont{Cao}},
  \bibinfo{journal}{Phys. Rev. B} \textbf{\bibinfo{volume}{96}},
  \bibinfo{pages}{064436} (\bibinfo{year}{2017}),
  \urlprefix\url{https://link.aps.org/doi/10.1103/PhysRevB.96.064436}.

\bibitem[{\citenamefont{Narayanan et~al.}(2010)\citenamefont{Narayanan,
  Mikhailova, Senyshyn, Trots, Laskowski, Blaha, Schwarz, Fuess, and
  Ehrenberg}}]{Narayanan.2010}
\bibinfo{author}{\bibfnamefont{N.}~\bibnamefont{Narayanan}},
  \bibinfo{author}{\bibfnamefont{D.}~\bibnamefont{Mikhailova}},
  \bibinfo{author}{\bibfnamefont{A.}~\bibnamefont{Senyshyn}},
  \bibinfo{author}{\bibfnamefont{D.~M.} \bibnamefont{Trots}},
  \bibinfo{author}{\bibfnamefont{R.}~\bibnamefont{Laskowski}},
  \bibinfo{author}{\bibfnamefont{P.}~\bibnamefont{Blaha}},
  \bibinfo{author}{\bibfnamefont{K.}~\bibnamefont{Schwarz}},
  \bibinfo{author}{\bibfnamefont{H.}~\bibnamefont{Fuess}}, \bibnamefont{and}
  \bibinfo{author}{\bibfnamefont{H.}~\bibnamefont{Ehrenberg}},
  \bibinfo{journal}{Phys. Rev. B} \textbf{\bibinfo{volume}{82}},
  \bibinfo{pages}{024403} (\bibinfo{year}{2010}),
  \urlprefix\url{https://link.aps.org/doi/10.1103/PhysRevB.82.024403}.

\bibitem[{\citenamefont{Mikhailova et~al.}(2010)\citenamefont{Mikhailova,
  Narayanan, Gruner, Voss, Senyshyn, Trots, Fuess, and
  Ehrenberg}}]{Mikhailova.2010}
\bibinfo{author}{\bibfnamefont{D.}~\bibnamefont{Mikhailova}},
  \bibinfo{author}{\bibfnamefont{N.}~\bibnamefont{Narayanan}},
  \bibinfo{author}{\bibfnamefont{W.}~\bibnamefont{Gruner}},
  \bibinfo{author}{\bibfnamefont{A.}~\bibnamefont{Voss}},
  \bibinfo{author}{\bibfnamefont{A.}~\bibnamefont{Senyshyn}},
  \bibinfo{author}{\bibfnamefont{D.~M.} \bibnamefont{Trots}},
  \bibinfo{author}{\bibfnamefont{H.}~\bibnamefont{Fuess}}, \bibnamefont{and}
  \bibinfo{author}{\bibfnamefont{H.}~\bibnamefont{Ehrenberg}},
  \bibinfo{journal}{Inorganic Chemistry} \textbf{\bibinfo{volume}{49}},
  \bibinfo{pages}{10348} (\bibinfo{year}{2010}), \bibinfo{note}{pMID:
  20964307}, \urlprefix\url{https://doi.org/10.1021/ic100973p}.

\bibitem[{\citenamefont{Kolchinskaya et~al.}(2012)\citenamefont{Kolchinskaya,
  Komissinskiy, Yazdi, Vafaee, Mikhailova, Narayanan, Ehrenberg, Wilhelm,
  Rogalev, and Alff}}]{Kolchinskaya.2012}
\bibinfo{author}{\bibfnamefont{A.}~\bibnamefont{Kolchinskaya}},
  \bibinfo{author}{\bibfnamefont{P.}~\bibnamefont{Komissinskiy}},
  \bibinfo{author}{\bibfnamefont{M.~B.} \bibnamefont{Yazdi}},
  \bibinfo{author}{\bibfnamefont{M.}~\bibnamefont{Vafaee}},
  \bibinfo{author}{\bibfnamefont{D.}~\bibnamefont{Mikhailova}},
  \bibinfo{author}{\bibfnamefont{N.}~\bibnamefont{Narayanan}},
  \bibinfo{author}{\bibfnamefont{H.}~\bibnamefont{Ehrenberg}},
  \bibinfo{author}{\bibfnamefont{F.}~\bibnamefont{Wilhelm}},
  \bibinfo{author}{\bibfnamefont{A.}~\bibnamefont{Rogalev}}, \bibnamefont{and}
  \bibinfo{author}{\bibfnamefont{L.}~\bibnamefont{Alff}},
  \bibinfo{journal}{Phys. Rev. B} \textbf{\bibinfo{volume}{85}},
  \bibinfo{pages}{224422} (\bibinfo{year}{2012}),
  \urlprefix\url{https://link.aps.org/doi/10.1103/PhysRevB.85.224422}.

\bibitem[{\citenamefont{Laguna-Marco et~al.}(2015)\citenamefont{Laguna-Marco,
  Kayser, Alonso, Mart\'{\i}nez-Lope, van Veenendaal, Choi, and
  Haskel}}]{Laguna.2015}
\bibinfo{author}{\bibfnamefont{M.~A.} \bibnamefont{Laguna-Marco}},
  \bibinfo{author}{\bibfnamefont{P.}~\bibnamefont{Kayser}},
  \bibinfo{author}{\bibfnamefont{J.~A.} \bibnamefont{Alonso}},
  \bibinfo{author}{\bibfnamefont{M.~J.} \bibnamefont{Mart\'{\i}nez-Lope}},
  \bibinfo{author}{\bibfnamefont{M.}~\bibnamefont{van Veenendaal}},
  \bibinfo{author}{\bibfnamefont{Y.}~\bibnamefont{Choi}}, \bibnamefont{and}
  \bibinfo{author}{\bibfnamefont{D.}~\bibnamefont{Haskel}},
  \bibinfo{journal}{Phys. Rev. B} \textbf{\bibinfo{volume}{91}},
  \bibinfo{pages}{214433} (\bibinfo{year}{2015}),
  \urlprefix\url{https://link.aps.org/doi/10.1103/PhysRevB.91.214433}.

\bibitem[{\citenamefont{Ohresser et~al.}(2014)\citenamefont{Ohresser, Otero,
  Choueikani, Chen, Stanescu, Deschamps, Moreno, Polack, Lagarde, Daguerre
  et~al.}}]{Ohresser.2014}
\bibinfo{author}{\bibfnamefont{P.}~\bibnamefont{Ohresser}},
  \bibinfo{author}{\bibfnamefont{E.}~\bibnamefont{Otero}},
  \bibinfo{author}{\bibfnamefont{F.}~\bibnamefont{Choueikani}},
  \bibinfo{author}{\bibfnamefont{K.}~\bibnamefont{Chen}},
  \bibinfo{author}{\bibfnamefont{S.}~\bibnamefont{Stanescu}},
  \bibinfo{author}{\bibfnamefont{F.}~\bibnamefont{Deschamps}},
  \bibinfo{author}{\bibfnamefont{T.}~\bibnamefont{Moreno}},
  \bibinfo{author}{\bibfnamefont{F.}~\bibnamefont{Polack}},
  \bibinfo{author}{\bibfnamefont{B.}~\bibnamefont{Lagarde}},
  \bibinfo{author}{\bibfnamefont{J.-P.} \bibnamefont{Daguerre}},
  \bibnamefont{et~al.}, \bibinfo{journal}{Review of Scientific Instruments}
  \textbf{\bibinfo{volume}{85}}, \bibinfo{pages}{013106}
  (\bibinfo{year}{2014}), \urlprefix\url{https://doi.org/10.1063/1.4861191}.

\bibitem[{\citenamefont{Rogalev and Wilhelm}(2015)}]{Rogalev.2015}
\bibinfo{author}{\bibfnamefont{A.}~\bibnamefont{Rogalev}} \bibnamefont{and}
  \bibinfo{author}{\bibfnamefont{F.}~\bibnamefont{Wilhelm}},
  \bibinfo{journal}{The Physics of Metals and Metallography}
  \textbf{\bibinfo{volume}{116}}, \bibinfo{pages}{1285} (\bibinfo{year}{2015}),
  ISSN \bibinfo{issn}{1555-6190},
  \urlprefix\url{https://doi.org/10.1134/S0031918X15130013}.

\bibitem[{\citenamefont{Henke et~al.}(1993)\citenamefont{Henke, Gullikson, and
  Davis}}]{Henke.1993}
\bibinfo{author}{\bibfnamefont{B.}~\bibnamefont{Henke}},
  \bibinfo{author}{\bibfnamefont{E.}~\bibnamefont{Gullikson}},
  \bibnamefont{and} \bibinfo{author}{\bibfnamefont{J.}~\bibnamefont{Davis}},
  \bibinfo{journal}{Atomic Data and Nuclear Data Tables}
  \textbf{\bibinfo{volume}{54}}, \bibinfo{pages}{181 } (\bibinfo{year}{1993}),
  ISSN \bibinfo{issn}{0092-640X},
  \urlprefix\url{http://www.sciencedirect.com/science/article/pii/S0092640X83710132}.

\bibitem[{\citenamefont{Haverkort et~al.}(2012)\citenamefont{Haverkort,
  Zwierzycki, and Andersen}}]{Haverkort.2012}
\bibinfo{author}{\bibfnamefont{M.~W.} \bibnamefont{Haverkort}},
  \bibinfo{author}{\bibfnamefont{M.}~\bibnamefont{Zwierzycki}},
  \bibnamefont{and} \bibinfo{author}{\bibfnamefont{O.~K.}
  \bibnamefont{Andersen}}, \bibinfo{journal}{Phys. Rev. B}
  \textbf{\bibinfo{volume}{85}}, \bibinfo{pages}{165113}
  (\bibinfo{year}{2012}),
  \urlprefix\url{https://link.aps.org/doi/10.1103/PhysRevB.85.165113}.

\bibitem[{\citenamefont{Lu et~al.}(2014)\citenamefont{Lu, H\"oppner,
  Gunnarsson, and Haverkort}}]{Lu.2014}
\bibinfo{author}{\bibfnamefont{Y.}~\bibnamefont{Lu}},
  \bibinfo{author}{\bibfnamefont{M.}~\bibnamefont{H\"oppner}},
  \bibinfo{author}{\bibfnamefont{O.}~\bibnamefont{Gunnarsson}},
  \bibnamefont{and} \bibinfo{author}{\bibfnamefont{M.~W.}
  \bibnamefont{Haverkort}}, \bibinfo{journal}{Phys. Rev. B}
  \textbf{\bibinfo{volume}{90}}, \bibinfo{pages}{085102}
  (\bibinfo{year}{2014}),
  \urlprefix\url{https://link.aps.org/doi/10.1103/PhysRevB.90.085102}.

\bibitem[{\citenamefont{Haverkort et~al.}(2014)\citenamefont{Haverkort,
  Sangiovanni, Hansmann, Toschi, Lu, and Macke}}]{Haverkort.2014}
\bibinfo{author}{\bibfnamefont{M.~W.} \bibnamefont{Haverkort}},
  \bibinfo{author}{\bibfnamefont{G.}~\bibnamefont{Sangiovanni}},
  \bibinfo{author}{\bibfnamefont{P.}~\bibnamefont{Hansmann}},
  \bibinfo{author}{\bibfnamefont{A.}~\bibnamefont{Toschi}},
  \bibinfo{author}{\bibfnamefont{Y.}~\bibnamefont{Lu}}, \bibnamefont{and}
  \bibinfo{author}{\bibfnamefont{S.}~\bibnamefont{Macke}},
  \bibinfo{journal}{EPL (Europhysics Letters)} \textbf{\bibinfo{volume}{108}},
  \bibinfo{pages}{57004} (\bibinfo{year}{2014}),
  \urlprefix\url{http://stacks.iop.org/0295-5075/108/i=5/a=57004}.

\bibitem[{\citenamefont{Koepernik and Eschrig}(1999)}]{Koepernik.1999}
\bibinfo{author}{\bibfnamefont{K.}~\bibnamefont{Koepernik}} \bibnamefont{and}
  \bibinfo{author}{\bibfnamefont{H.}~\bibnamefont{Eschrig}},
  \bibinfo{journal}{Phys. Rev. B} \textbf{\bibinfo{volume}{59}},
  \bibinfo{pages}{1743} (\bibinfo{year}{1999}),
  \urlprefix\url{https://link.aps.org/doi/10.1103/PhysRevB.59.1743}.

\bibitem[{Cop()}]{Coparameters.2019}
\bibinfo{note}{{CoO6 cluster parameters [eV]: $U_{dd}=5.5$, $U_{pd}=7.0$, ionic
  crystal field 10$D_q=0.545$, $\Delta t_{2g}= -0.265$, $\Delta e_g$= -0.047,
  charge transfer energy $\Delta_{CT}= 2.0$, hybridization $V(x^2-y^2) = 2.52$,
  $V(z^2) = 2.63$, $V(xy) = 1.38$, $V(xz) = V(yz) = 1.43$, ligand-crystal field
  10$Dq^{lig} = 1.0$, spin-orbit coupling $ \zeta_{3d}=0.074$, exchange field
  $H_{ex} = 0.012$ and magnetic field 6 T. The Slater integrals were reduced to
  80\% of Hartree-Fock values. In the calculations of the effective moment at
  high temperature the exchange field was considered to be zero.}}

\bibitem[{Irp()}]{Irparameters.2019}
\bibinfo{note}{{IrO6 cluster parameters [eV]: $U_{dd}=1.0$, $U_{pd}=2.0$, ionic
  crystal field 10$D_q=2.3$, $\Delta t_{2g}= 0.1$, $\Delta e_g$= 0.125, charge
  transfer energy $\Delta_{CT}= -1.5$, hybridization $V(x^2-y^2) = 5.21$,
  $V(z^2) = 5.08$, $V(xy) = 2.835$, $V(xz) = V(yz) = 2.915$, ligand-crystal
  field 10$Dq^{lig} = 0.93$, spin-orbit coupling $ \zeta_{5d}=0.4$, exchange
  field $H_{ex} = 0.002$ and magnetic field 17 T. The Slater integrals were
  reduced to 70\% of Hartree-Fock values. In the calculations of the effective
  moment at high temperature the exchange field was considered to be zero.}}

\bibitem[{\citenamefont{Wong et~al.}(1984)\citenamefont{Wong, Lytle, Messmer,
  and Maylotte}}]{Wong.1984}
\bibinfo{author}{\bibfnamefont{J.}~\bibnamefont{Wong}},
  \bibinfo{author}{\bibfnamefont{F.~W.} \bibnamefont{Lytle}},
  \bibinfo{author}{\bibfnamefont{R.~P.} \bibnamefont{Messmer}},
  \bibnamefont{and} \bibinfo{author}{\bibfnamefont{D.~H.}
  \bibnamefont{Maylotte}}, \bibinfo{journal}{Phys. Rev. B}
  \textbf{\bibinfo{volume}{30}}, \bibinfo{pages}{5596} (\bibinfo{year}{1984}),
  \urlprefix\url{https://link.aps.org/doi/10.1103/PhysRevB.30.5596}.

\bibitem[{\citenamefont{Chen and Sette}(1990)}]{Chen.1990}
\bibinfo{author}{\bibfnamefont{C.~T.} \bibnamefont{Chen}} \bibnamefont{and}
  \bibinfo{author}{\bibfnamefont{F.}~\bibnamefont{Sette}},
  \bibinfo{journal}{Physica Scripta} \textbf{\bibinfo{volume}{1990}},
  \bibinfo{pages}{119} (\bibinfo{year}{1990}),
  \urlprefix\url{http://stacks.iop.org/1402-4896/1990/i=T31/a=016}.

\bibitem[{\citenamefont{Mitra et~al.}(2003)\citenamefont{Mitra, Hu,
  Raychaudhuri, Wirth, Csiszar, Hsieh, Lin, Chen, and Tjeng}}]{Mitra.2003}
\bibinfo{author}{\bibfnamefont{C.}~\bibnamefont{Mitra}},
  \bibinfo{author}{\bibfnamefont{Z.}~\bibnamefont{Hu}},
  \bibinfo{author}{\bibfnamefont{P.}~\bibnamefont{Raychaudhuri}},
  \bibinfo{author}{\bibfnamefont{S.}~\bibnamefont{Wirth}},
  \bibinfo{author}{\bibfnamefont{S.~I.} \bibnamefont{Csiszar}},
  \bibinfo{author}{\bibfnamefont{H.~H.} \bibnamefont{Hsieh}},
  \bibinfo{author}{\bibfnamefont{H.-J.} \bibnamefont{Lin}},
  \bibinfo{author}{\bibfnamefont{C.~T.} \bibnamefont{Chen}}, \bibnamefont{and}
  \bibinfo{author}{\bibfnamefont{L.~H.} \bibnamefont{Tjeng}},
  \bibinfo{journal}{Phys. Rev. B} \textbf{\bibinfo{volume}{67}},
  \bibinfo{pages}{092404} (\bibinfo{year}{2003}),
  \urlprefix\url{https://link.aps.org/doi/10.1103/PhysRevB.67.092404}.

\bibitem[{\citenamefont{Burnus et~al.}(2006)\citenamefont{Burnus, Hu,
  Haverkort, Cezar, Flahaut, Hardy, Maignan, Brookes, Tanaka, Hsieh
  et~al.}}]{Burnus.2006}
\bibinfo{author}{\bibfnamefont{T.}~\bibnamefont{Burnus}},
  \bibinfo{author}{\bibfnamefont{Z.}~\bibnamefont{Hu}},
  \bibinfo{author}{\bibfnamefont{M.~W.} \bibnamefont{Haverkort}},
  \bibinfo{author}{\bibfnamefont{J.~C.} \bibnamefont{Cezar}},
  \bibinfo{author}{\bibfnamefont{D.}~\bibnamefont{Flahaut}},
  \bibinfo{author}{\bibfnamefont{V.}~\bibnamefont{Hardy}},
  \bibinfo{author}{\bibfnamefont{A.}~\bibnamefont{Maignan}},
  \bibinfo{author}{\bibfnamefont{N.~B.} \bibnamefont{Brookes}},
  \bibinfo{author}{\bibfnamefont{A.}~\bibnamefont{Tanaka}},
  \bibinfo{author}{\bibfnamefont{H.~H.} \bibnamefont{Hsieh}},
  \bibnamefont{et~al.}, \bibinfo{journal}{Phys. Rev. B}
  \textbf{\bibinfo{volume}{74}}, \bibinfo{pages}{245111}
  (\bibinfo{year}{2006}),
  \urlprefix\url{https://link.aps.org/doi/10.1103/PhysRevB.74.245111}.

\bibitem[{\citenamefont{Burnus et~al.}(2008{\natexlab{a}})\citenamefont{Burnus,
  Hu, Hsieh, Joly, Joy, Haverkort, Wu, Tanaka, Lin, Chen et~al.}}]{Burnus.2008}
\bibinfo{author}{\bibfnamefont{T.}~\bibnamefont{Burnus}},
  \bibinfo{author}{\bibfnamefont{Z.}~\bibnamefont{Hu}},
  \bibinfo{author}{\bibfnamefont{H.~H.} \bibnamefont{Hsieh}},
  \bibinfo{author}{\bibfnamefont{V.~L.~J.} \bibnamefont{Joly}},
  \bibinfo{author}{\bibfnamefont{P.~A.} \bibnamefont{Joy}},
  \bibinfo{author}{\bibfnamefont{M.~W.} \bibnamefont{Haverkort}},
  \bibinfo{author}{\bibfnamefont{H.}~\bibnamefont{Wu}},
  \bibinfo{author}{\bibfnamefont{A.}~\bibnamefont{Tanaka}},
  \bibinfo{author}{\bibfnamefont{H.-J.} \bibnamefont{Lin}},
  \bibinfo{author}{\bibfnamefont{C.~T.} \bibnamefont{Chen}},
  \bibnamefont{et~al.}, \bibinfo{journal}{Phys. Rev. B}
  \textbf{\bibinfo{volume}{77}}, \bibinfo{pages}{125124}
  (\bibinfo{year}{2008}{\natexlab{a}}),
  \urlprefix\url{https://link.aps.org/doi/10.1103/PhysRevB.77.125124}.

\bibitem[{\citenamefont{Baroudi et~al.}(2014)\citenamefont{Baroudi, Yim, Wu,
  Huang, Roudebush, Vavilova, Grafe, Kataev, Buechner, Ji
  et~al.}}]{Baroudi.2014}
\bibinfo{author}{\bibfnamefont{K.}~\bibnamefont{Baroudi}},
  \bibinfo{author}{\bibfnamefont{C.}~\bibnamefont{Yim}},
  \bibinfo{author}{\bibfnamefont{H.}~\bibnamefont{Wu}},
  \bibinfo{author}{\bibfnamefont{Q.}~\bibnamefont{Huang}},
  \bibinfo{author}{\bibfnamefont{J.~H.} \bibnamefont{Roudebush}},
  \bibinfo{author}{\bibfnamefont{E.}~\bibnamefont{Vavilova}},
  \bibinfo{author}{\bibfnamefont{H.-J.} \bibnamefont{Grafe}},
  \bibinfo{author}{\bibfnamefont{V.}~\bibnamefont{Kataev}},
  \bibinfo{author}{\bibfnamefont{B.}~\bibnamefont{Buechner}},
  \bibinfo{author}{\bibfnamefont{H.}~\bibnamefont{Ji}}, \bibnamefont{et~al.},
  \bibinfo{journal}{Journal of Solid State Chemistry}
  \textbf{\bibinfo{volume}{210}}, \bibinfo{pages}{195 } (\bibinfo{year}{2014}),
  ISSN \bibinfo{issn}{0022-4596},
  \urlprefix\url{http://www.sciencedirect.com/science/article/pii/S002245961300563X}.

\bibitem[{\citenamefont{Choy et~al.}(1995)\citenamefont{Choy, Kim, Hwang,
  Demazeau, and Jung}}]{Choy.1995}
\bibinfo{author}{\bibfnamefont{J.-H.} \bibnamefont{Choy}},
  \bibinfo{author}{\bibfnamefont{D.-K.} \bibnamefont{Kim}},
  \bibinfo{author}{\bibfnamefont{S.-H.} \bibnamefont{Hwang}},
  \bibinfo{author}{\bibfnamefont{G.}~\bibnamefont{Demazeau}}, \bibnamefont{and}
  \bibinfo{author}{\bibfnamefont{D.-Y.} \bibnamefont{Jung}},
  \bibinfo{journal}{Journal of the American Chemical Society}
  \textbf{\bibinfo{volume}{117}}, \bibinfo{pages}{8557} (\bibinfo{year}{1995}),
  \urlprefix\url{https://doi.org/10.1021/ja00138a010}.

\bibitem[{\citenamefont{Mugavero~III et~al.}(2009)\citenamefont{Mugavero~III,
  Smith, Yoon, and zur Loye}}]{Mugavero.2009}
\bibinfo{author}{\bibfnamefont{S.}~\bibnamefont{Mugavero~III}},
  \bibinfo{author}{\bibfnamefont{M.}~\bibnamefont{Smith}},
  \bibinfo{author}{\bibfnamefont{W.-S.} \bibnamefont{Yoon}}, \bibnamefont{and}
  \bibinfo{author}{\bibfnamefont{H.-C.} \bibnamefont{zur Loye}},
  \bibinfo{journal}{Angewandte Chemie International Edition}
  \textbf{\bibinfo{volume}{48}}, \bibinfo{pages}{215} (\bibinfo{year}{2009}),
  \urlprefix\url{https://onlinelibrary.wiley.com/doi/abs/10.1002/anie.200804045}.

\bibitem[{\citenamefont{Chen et~al.}(2014)\citenamefont{Chen, Chin, Valldor,
  Hu, Lee, Haw, Hiraoka, Ishii, Pao, Tsuei et~al.}}]{Chen.2014}
\bibinfo{author}{\bibfnamefont{J.-M.} \bibnamefont{Chen}},
  \bibinfo{author}{\bibfnamefont{Y.-Y.} \bibnamefont{Chin}},
  \bibinfo{author}{\bibfnamefont{M.}~\bibnamefont{Valldor}},
  \bibinfo{author}{\bibfnamefont{Z.}~\bibnamefont{Hu}},
  \bibinfo{author}{\bibfnamefont{J.-M.} \bibnamefont{Lee}},
  \bibinfo{author}{\bibfnamefont{S.-C.} \bibnamefont{Haw}},
  \bibinfo{author}{\bibfnamefont{N.}~\bibnamefont{Hiraoka}},
  \bibinfo{author}{\bibfnamefont{H.}~\bibnamefont{Ishii}},
  \bibinfo{author}{\bibfnamefont{C.-W.} \bibnamefont{Pao}},
  \bibinfo{author}{\bibfnamefont{K.-D.} \bibnamefont{Tsuei}},
  \bibnamefont{et~al.}, \bibinfo{journal}{Journal of the American Chemical
  Society} \textbf{\bibinfo{volume}{136}}, \bibinfo{pages}{1514}
  (\bibinfo{year}{2014}), \bibinfo{note}{pMID: 24410074},
  \urlprefix\url{https://doi.org/10.1021/ja4114006}.

\bibitem[{\citenamefont{Esser et~al.}(2018)\citenamefont{Esser, Chang, Kuo,
  Merten, Roddatis, Ha, Jesche, Moshnyaga, Lin, Tanaka et~al.}}]{Esser.2018}
\bibinfo{author}{\bibfnamefont{S.}~\bibnamefont{Esser}},
  \bibinfo{author}{\bibfnamefont{C.~F.} \bibnamefont{Chang}},
  \bibinfo{author}{\bibfnamefont{C.-Y.} \bibnamefont{Kuo}},
  \bibinfo{author}{\bibfnamefont{S.}~\bibnamefont{Merten}},
  \bibinfo{author}{\bibfnamefont{V.}~\bibnamefont{Roddatis}},
  \bibinfo{author}{\bibfnamefont{T.~D.} \bibnamefont{Ha}},
  \bibinfo{author}{\bibfnamefont{A.}~\bibnamefont{Jesche}},
  \bibinfo{author}{\bibfnamefont{V.}~\bibnamefont{Moshnyaga}},
  \bibinfo{author}{\bibfnamefont{H.-J.} \bibnamefont{Lin}},
  \bibinfo{author}{\bibfnamefont{A.}~\bibnamefont{Tanaka}},
  \bibnamefont{et~al.}, \bibinfo{journal}{Phys. Rev. B}
  \textbf{\bibinfo{volume}{97}}, \bibinfo{pages}{205121}
  (\bibinfo{year}{2018}),
  \urlprefix\url{https://link.aps.org/doi/10.1103/PhysRevB.97.205121}.

\bibitem[{\citenamefont{Hu et~al.}(2004)\citenamefont{Hu, Wu, Haverkort, Hsieh,
  Lin, Lorenz, Baier, Reichl, Bonn, Felser et~al.}}]{Hu.2004}
\bibinfo{author}{\bibfnamefont{Z.}~\bibnamefont{Hu}},
  \bibinfo{author}{\bibfnamefont{H.}~\bibnamefont{Wu}},
  \bibinfo{author}{\bibfnamefont{M.~W.} \bibnamefont{Haverkort}},
  \bibinfo{author}{\bibfnamefont{H.~H.} \bibnamefont{Hsieh}},
  \bibinfo{author}{\bibfnamefont{H.~J.} \bibnamefont{Lin}},
  \bibinfo{author}{\bibfnamefont{T.}~\bibnamefont{Lorenz}},
  \bibinfo{author}{\bibfnamefont{J.}~\bibnamefont{Baier}},
  \bibinfo{author}{\bibfnamefont{A.}~\bibnamefont{Reichl}},
  \bibinfo{author}{\bibfnamefont{I.}~\bibnamefont{Bonn}},
  \bibinfo{author}{\bibfnamefont{C.}~\bibnamefont{Felser}},
  \bibnamefont{et~al.}, \bibinfo{journal}{Phys. Rev. Lett.}
  \textbf{\bibinfo{volume}{92}}, \bibinfo{pages}{207402}
  (\bibinfo{year}{2004}),
  \urlprefix\url{https://link.aps.org/doi/10.1103/PhysRevLett.92.207402}.

\bibitem[{\citenamefont{Haverkort et~al.}(2006)\citenamefont{Haverkort, Hu,
  Cezar, Burnus, Hartmann, Reuther, Zobel, Lorenz, Tanaka, Brookes
  et~al.}}]{Haverkort.2006}
\bibinfo{author}{\bibfnamefont{M.~W.} \bibnamefont{Haverkort}},
  \bibinfo{author}{\bibfnamefont{Z.}~\bibnamefont{Hu}},
  \bibinfo{author}{\bibfnamefont{J.~C.} \bibnamefont{Cezar}},
  \bibinfo{author}{\bibfnamefont{T.}~\bibnamefont{Burnus}},
  \bibinfo{author}{\bibfnamefont{H.}~\bibnamefont{Hartmann}},
  \bibinfo{author}{\bibfnamefont{M.}~\bibnamefont{Reuther}},
  \bibinfo{author}{\bibfnamefont{C.}~\bibnamefont{Zobel}},
  \bibinfo{author}{\bibfnamefont{T.}~\bibnamefont{Lorenz}},
  \bibinfo{author}{\bibfnamefont{A.}~\bibnamefont{Tanaka}},
  \bibinfo{author}{\bibfnamefont{N.~B.} \bibnamefont{Brookes}},
  \bibnamefont{et~al.}, \bibinfo{journal}{Phys. Rev. Lett.}
  \textbf{\bibinfo{volume}{97}}, \bibinfo{pages}{176405}
  (\bibinfo{year}{2006}),
  \urlprefix\url{https://link.aps.org/doi/10.1103/PhysRevLett.97.176405}.

\bibitem[{\citenamefont{Chang et~al.}(2009)\citenamefont{Chang, Hu, Wu, Burnus,
  Hollmann, Benomar, Lorenz, Tanaka, Lin, Hsieh et~al.}}]{Chang.2009}
\bibinfo{author}{\bibfnamefont{C.~F.} \bibnamefont{Chang}},
  \bibinfo{author}{\bibfnamefont{Z.}~\bibnamefont{Hu}},
  \bibinfo{author}{\bibfnamefont{H.}~\bibnamefont{Wu}},
  \bibinfo{author}{\bibfnamefont{T.}~\bibnamefont{Burnus}},
  \bibinfo{author}{\bibfnamefont{N.}~\bibnamefont{Hollmann}},
  \bibinfo{author}{\bibfnamefont{M.}~\bibnamefont{Benomar}},
  \bibinfo{author}{\bibfnamefont{T.}~\bibnamefont{Lorenz}},
  \bibinfo{author}{\bibfnamefont{A.}~\bibnamefont{Tanaka}},
  \bibinfo{author}{\bibfnamefont{H.-J.} \bibnamefont{Lin}},
  \bibinfo{author}{\bibfnamefont{H.~H.} \bibnamefont{Hsieh}},
  \bibnamefont{et~al.}, \bibinfo{journal}{Phys. Rev. Lett.}
  \textbf{\bibinfo{volume}{102}}, \bibinfo{pages}{116401}
  (\bibinfo{year}{2009}),
  \urlprefix\url{https://link.aps.org/doi/10.1103/PhysRevLett.102.116401}.

\bibitem[{\citenamefont{Burnus et~al.}(2008{\natexlab{b}})\citenamefont{Burnus,
  Hu, Wu, Cezar, Niitaka, Takagi, Chang, Brookes, Lin, Jang
  et~al.}}]{Burnus.2008b}
\bibinfo{author}{\bibfnamefont{T.}~\bibnamefont{Burnus}},
  \bibinfo{author}{\bibfnamefont{Z.}~\bibnamefont{Hu}},
  \bibinfo{author}{\bibfnamefont{H.}~\bibnamefont{Wu}},
  \bibinfo{author}{\bibfnamefont{J.~C.} \bibnamefont{Cezar}},
  \bibinfo{author}{\bibfnamefont{S.}~\bibnamefont{Niitaka}},
  \bibinfo{author}{\bibfnamefont{H.}~\bibnamefont{Takagi}},
  \bibinfo{author}{\bibfnamefont{C.~F.} \bibnamefont{Chang}},
  \bibinfo{author}{\bibfnamefont{N.~B.} \bibnamefont{Brookes}},
  \bibinfo{author}{\bibfnamefont{H.-J.} \bibnamefont{Lin}},
  \bibinfo{author}{\bibfnamefont{L.~Y.} \bibnamefont{Jang}},
  \bibnamefont{et~al.}, \bibinfo{journal}{Phys. Rev. B}
  \textbf{\bibinfo{volume}{77}}, \bibinfo{pages}{205111}
  (\bibinfo{year}{2008}{\natexlab{b}}),
  \urlprefix\url{https://link.aps.org/doi/10.1103/PhysRevB.77.205111}.

\bibitem[{\citenamefont{Burnus et~al.}(2008{\natexlab{c}})\citenamefont{Burnus,
  Hu, Wu, Cezar, Niitaka, Takagi, Chang, Brookes, Lin, Jang
  et~al.}}]{Burnus.2008c}
\bibinfo{author}{\bibfnamefont{T.}~\bibnamefont{Burnus}},
  \bibinfo{author}{\bibfnamefont{Z.}~\bibnamefont{Hu}},
  \bibinfo{author}{\bibfnamefont{H.}~\bibnamefont{Wu}},
  \bibinfo{author}{\bibfnamefont{J.~C.} \bibnamefont{Cezar}},
  \bibinfo{author}{\bibfnamefont{S.}~\bibnamefont{Niitaka}},
  \bibinfo{author}{\bibfnamefont{H.}~\bibnamefont{Takagi}},
  \bibinfo{author}{\bibfnamefont{C.~F.} \bibnamefont{Chang}},
  \bibinfo{author}{\bibfnamefont{N.~B.} \bibnamefont{Brookes}},
  \bibinfo{author}{\bibfnamefont{H.-J.} \bibnamefont{Lin}},
  \bibinfo{author}{\bibfnamefont{L.~Y.} \bibnamefont{Jang}},
  \bibnamefont{et~al.}, \bibinfo{journal}{Phys. Rev. B}
  \textbf{\bibinfo{volume}{77}}, \bibinfo{pages}{205111}
  (\bibinfo{year}{2008}{\natexlab{c}}),
  \urlprefix\url{https://link.aps.org/doi/10.1103/PhysRevB.77.205111}.

\bibitem[{\citenamefont{Hollmann et~al.}(2009)\citenamefont{Hollmann, Hu,
  Valldor, Maignan, Tanaka, Hsieh, Lin, Chen, and Tjeng}}]{Hollmann.2009}
\bibinfo{author}{\bibfnamefont{N.}~\bibnamefont{Hollmann}},
  \bibinfo{author}{\bibfnamefont{Z.}~\bibnamefont{Hu}},
  \bibinfo{author}{\bibfnamefont{M.}~\bibnamefont{Valldor}},
  \bibinfo{author}{\bibfnamefont{A.}~\bibnamefont{Maignan}},
  \bibinfo{author}{\bibfnamefont{A.}~\bibnamefont{Tanaka}},
  \bibinfo{author}{\bibfnamefont{H.~H.} \bibnamefont{Hsieh}},
  \bibinfo{author}{\bibfnamefont{H.-J.} \bibnamefont{Lin}},
  \bibinfo{author}{\bibfnamefont{C.~T.} \bibnamefont{Chen}}, \bibnamefont{and}
  \bibinfo{author}{\bibfnamefont{L.~H.} \bibnamefont{Tjeng}},
  \bibinfo{journal}{Phys. Rev. B} \textbf{\bibinfo{volume}{80}},
  \bibinfo{pages}{085111} (\bibinfo{year}{2009}),
  \urlprefix\url{https://link.aps.org/doi/10.1103/PhysRevB.80.085111}.

\bibitem[{\citenamefont{Thole et~al.}(1992)\citenamefont{Thole, Carra, Sette,
  and van~der Laan}}]{Thole.1992}
\bibinfo{author}{\bibfnamefont{B.~T.} \bibnamefont{Thole}},
  \bibinfo{author}{\bibfnamefont{P.}~\bibnamefont{Carra}},
  \bibinfo{author}{\bibfnamefont{F.}~\bibnamefont{Sette}}, \bibnamefont{and}
  \bibinfo{author}{\bibfnamefont{G.}~\bibnamefont{van~der Laan}},
  \bibinfo{journal}{Phys. Rev. Lett.} \textbf{\bibinfo{volume}{68}},
  \bibinfo{pages}{1943} (\bibinfo{year}{1992}),
  \urlprefix\url{https://link.aps.org/doi/10.1103/PhysRevLett.68.1943}.

\bibitem[{\citenamefont{Carra et~al.}(1993)\citenamefont{Carra, Thole,
  Altarelli, and Wang}}]{Carra.1993}
\bibinfo{author}{\bibfnamefont{P.}~\bibnamefont{Carra}},
  \bibinfo{author}{\bibfnamefont{B.~T.} \bibnamefont{Thole}},
  \bibinfo{author}{\bibfnamefont{M.}~\bibnamefont{Altarelli}},
  \bibnamefont{and} \bibinfo{author}{\bibfnamefont{X.}~\bibnamefont{Wang}},
  \bibinfo{journal}{Phys. Rev. Lett.} \textbf{\bibinfo{volume}{70}},
  \bibinfo{pages}{694} (\bibinfo{year}{1993}),
  \urlprefix\url{https://link.aps.org/doi/10.1103/PhysRevLett.70.694}.

\bibitem[{\citenamefont{Teramura et~al.}(1996)\citenamefont{Teramura, Tanaka,
  and Jo}}]{Teramura.1996}
\bibinfo{author}{\bibfnamefont{Y.}~\bibnamefont{Teramura}},
  \bibinfo{author}{\bibfnamefont{A.}~\bibnamefont{Tanaka}}, \bibnamefont{and}
  \bibinfo{author}{\bibfnamefont{T.}~\bibnamefont{Jo}},
  \bibinfo{journal}{Journal of the Physical Society of Japan}
  \textbf{\bibinfo{volume}{65}}, \bibinfo{pages}{1053} (\bibinfo{year}{1996}),
  \urlprefix\url{https://doi.org/10.1143/JPSJ.65.1053}.

\bibitem[{\citenamefont{Laguna-Marco et~al.}(2010)\citenamefont{Laguna-Marco,
  Haskel, Souza-Neto, Lang, Krishnamurthy, Chikara, Cao, and van
  Veenendaal}}]{Laguna.2010}
\bibinfo{author}{\bibfnamefont{M.~A.} \bibnamefont{Laguna-Marco}},
  \bibinfo{author}{\bibfnamefont{D.}~\bibnamefont{Haskel}},
  \bibinfo{author}{\bibfnamefont{N.}~\bibnamefont{Souza-Neto}},
  \bibinfo{author}{\bibfnamefont{J.~C.} \bibnamefont{Lang}},
  \bibinfo{author}{\bibfnamefont{V.~V.} \bibnamefont{Krishnamurthy}},
  \bibinfo{author}{\bibfnamefont{S.}~\bibnamefont{Chikara}},
  \bibinfo{author}{\bibfnamefont{G.}~\bibnamefont{Cao}}, \bibnamefont{and}
  \bibinfo{author}{\bibfnamefont{M.}~\bibnamefont{van Veenendaal}},
  \bibinfo{journal}{Phys. Rev. Lett.} \textbf{\bibinfo{volume}{105}},
  \bibinfo{pages}{216407} (\bibinfo{year}{2010}),
  \urlprefix\url{https://link.aps.org/doi/10.1103/PhysRevLett.105.216407}.

\bibitem[{\citenamefont{Haskel et~al.}(2012)\citenamefont{Haskel, Fabbris,
  Zhernenkov, Kong, Jin, Cao, and van Veenendaal}}]{Haskel.2012}
\bibinfo{author}{\bibfnamefont{D.}~\bibnamefont{Haskel}},
  \bibinfo{author}{\bibfnamefont{G.}~\bibnamefont{Fabbris}},
  \bibinfo{author}{\bibfnamefont{M.}~\bibnamefont{Zhernenkov}},
  \bibinfo{author}{\bibfnamefont{P.~P.} \bibnamefont{Kong}},
  \bibinfo{author}{\bibfnamefont{C.~Q.} \bibnamefont{Jin}},
  \bibinfo{author}{\bibfnamefont{G.}~\bibnamefont{Cao}}, \bibnamefont{and}
  \bibinfo{author}{\bibfnamefont{M.}~\bibnamefont{van Veenendaal}},
  \bibinfo{journal}{Phys. Rev. Lett.} \textbf{\bibinfo{volume}{109}},
  \bibinfo{pages}{027204} (\bibinfo{year}{2012}),
  \urlprefix\url{https://link.aps.org/doi/10.1103/PhysRevLett.109.027204}.

\bibitem[{\citenamefont{Feng et~al.}(2016)\citenamefont{Feng, Yamaura, Tjeng,
  and Jansen}}]{Feng.2016}
\bibinfo{author}{\bibfnamefont{H.~L.} \bibnamefont{Feng}},
  \bibinfo{author}{\bibfnamefont{K.}~\bibnamefont{Yamaura}},
  \bibinfo{author}{\bibfnamefont{L.~H.} \bibnamefont{Tjeng}}, \bibnamefont{and}
  \bibinfo{author}{\bibfnamefont{M.}~\bibnamefont{Jansen}},
  \bibinfo{journal}{Journal of Solid State Chemistry}
  \textbf{\bibinfo{volume}{243}}, \bibinfo{pages}{119 } (\bibinfo{year}{2016}),
  ISSN \bibinfo{issn}{0022-4596},
  \urlprefix\url{http://www.sciencedirect.com/science/article/pii/S0022459616303310}.

\bibitem[{\citenamefont{Wolff et~al.}(2019)\citenamefont{Wolff, Tjeng, and
  Jansen}}]{Wolff.2019}
\bibinfo{author}{\bibfnamefont{K.~K.} \bibnamefont{Wolff}},
  \bibinfo{author}{\bibfnamefont{L.~H.} \bibnamefont{Tjeng}}, \bibnamefont{and}
  \bibinfo{author}{\bibfnamefont{M.}~\bibnamefont{Jansen}},
  \bibinfo{journal}{Solid State Communications} \textbf{\bibinfo{volume}{289}},
  \bibinfo{pages}{43 } (\bibinfo{year}{2019}), ISSN \bibinfo{issn}{0038-1098},
  \urlprefix\url{http://www.sciencedirect.com/science/article/pii/S0038109818306616}.

\bibitem[{\citenamefont{Bain and Berry}(2008)}]{Bain.2008}
\bibinfo{author}{\bibfnamefont{G.~A.} \bibnamefont{Bain}} \bibnamefont{and}
  \bibinfo{author}{\bibfnamefont{J.~F.} \bibnamefont{Berry}},
  \bibinfo{journal}{Journal of Chemical Education}
  \textbf{\bibinfo{volume}{85}}, \bibinfo{pages}{532} (\bibinfo{year}{2008}),
  \urlprefix\url{https://doi.org/10.1021/ed085p532}.

\bibitem[{\citenamefont{Wolff et~al.}(2017)\citenamefont{Wolff, Agrestini,
  Tanaka, Jansen, and Tjeng}}]{Wolff.2017}
\bibinfo{author}{\bibfnamefont{K.~K.} \bibnamefont{Wolff}},
  \bibinfo{author}{\bibfnamefont{S.}~\bibnamefont{Agrestini}},
  \bibinfo{author}{\bibfnamefont{A.}~\bibnamefont{Tanaka}},
  \bibinfo{author}{\bibfnamefont{M.}~\bibnamefont{Jansen}}, \bibnamefont{and}
  \bibinfo{author}{\bibfnamefont{L.~H.} \bibnamefont{Tjeng}},
  \bibinfo{journal}{Zeitschrift für anorganische und allgemeine Chemie}
  \textbf{\bibinfo{volume}{643}}, \bibinfo{pages}{2095} (\bibinfo{year}{2017}),
  \urlprefix\url{https://onlinelibrary.wiley.com/doi/abs/10.1002/zaac.201700386}.

\end{thebibliography}

\end{document}